\documentclass{fundam}

\publyear{25}
\papernumber{3}
\volume{194}
\issue{1}
\theDOI{10.46298/fi.11227}

\versionForARXIV

\usepackage{subcaption}

\usepackage{url} 
\usepackage[ruled,lined]{algorithm2e}
\usepackage{graphicx}
\usepackage{tikz}
\usetikzlibrary{automata, positioning, arrows}

\usepackage{amsfonts}
\usepackage{extarrows}
\usepackage{graphicx}
\usepackage{multirow}

\newcommand{\blank}{\hspace{0.6mm}\rule{2.3mm}{0.4pt}}
\newcommand{\intI}{\iota^\mathrm{I}}
\newcommand{\intII}{\iota^\mathrm{I\times U}}
\newcommand{\intIII}{\iota^\mathrm{N\times U}}
\newcommand{\evoI}{\epsilon^\mathrm{I}}
\newcommand{\evoII}{\epsilon^\mathrm{N}}

\definecolor{mygreen}{RGB}{0,150,45}
\definecolor{myyellow}{RGB}{150,130,00}

\theoremstyle{plain}
\THstyle

\newtheorem{notation}{Notation}[section]

\newcommand\myh{\parbox[0pt][5mm][c]{0cm}{}}

\begin{document}

\title{On the Computational Power of Particle Methods}

\address{Ivo Sbalzarini, Pfotenhauerstr.~108, 01307 Dresden, Germany; \url{sbalzarini@mpi-cbg.de}}

\author{Johannes Pahlke\\
Dresden University of Technology, Faculty of Computer Science, Dresden, Germany\\
Max Planck Institute of Molecular Cell Biology and Genetics, Dresden, Germany\\
Center for Systems Biology Dresden, Dresden Germany
\and Ivo F. Sbalzarini\\
Dresden University of Technology, Faculty of Computer Science, Dresden, Germany\\
Max Planck Institute of Molecular Cell Biology and Genetics, Dresden, Germany\\
Center for Systems Biology Dresden, Dresden Germany\\
Center for Scalable Data Analytics and Artificial Intelligence (ScaDS.AI) Dresden/Leipzig, Germany} 

\maketitle

\runninghead{Pahlke \& Sbalzarini}{Computational Power of Particle Methods}

\begin{abstract}
We investigate the computational power of particle methods, a well-established class of algorithms with applications in scientific computing and computer simulation. The computational power of a compute model determines the class of problems it can solve. Automata theory allows describing the computational power of abstract machines (automata) and the problems they can solve. At the top of the Chomsky hierarchy of formal languages and grammars are Turing machines, which resemble the concept on which most modern computers are built. Although particle methods can be interpreted as automata based on their formal definition, their computational power has so far not been studied. We address this by analyzing Turing completeness of particle methods. In particular, we prove two sets of restrictions under which a particle method is still Turing powerful, and we show when it loses Turing powerfulness. This contributes to understanding the theoretical foundations of particle methods and provides insight into the powerfulness of computer simulations. 
\end{abstract}

\begin{keywords}
particle methods, computational power, Turing completeness, halting decidability
\end{keywords}

\section{Introduction}
The computational power of an automaton is measured in the formal languages it can accept. 
This relates automata theory to formal languages and grammars, hence, to the Chomsky hierarchy \cite{Chomsky:1956}. 
According to this hierarchy and the Church-Turing hypothesis \cite{Turing1936,Church:1936}, the most powerful automaton is the Turing machine \cite{Turing1936}, albeit, to the best of our knowledge, there is no proof that a more powerful automaton cannot exist.
Turing machines resemble the concept on which most modern computers are built. 
Yet, a Turing machine is more powerful than any actual computer, since it has an infinitely long
tape (memory).
If an automaton is at least as powerful as a Turing machine, it is called \emph{Turing complete}.
Many systems are, in theory, Turing complete, including programming languages like Java, Python, C++, and even the C++ templates on their own \cite{Veldhuizen:2019}. But Turing completeness can also be found in more surprising systems, like Conway's game of life \cite{Rendell:2002,Rendell:2016}, the so-called ``Rule 110'' cellular automaton \cite{Cook:2004},  DNA with restriction enzymes \cite{Rothemund:1995}, and the card game ``Magic: The Gathering'' \cite{Churchill:2019}.
Proving Turing completeness of an automaton can help understand the theoretical capabilities and limitations of the system is models.
Here, we study Turing completeness of particle methods, a widely used class of algorithms in scientific computing and computer simulation.

Particle methods are a well-defined class of algorithms \cite{Pahlke:2023} with broad applicability. They are used, for example, to simulate fluid physics using the method of Smoothed Particle Hydrodynamics (SPH) \cite{Gingold:1977,Monaghan:2005,Lucy:1977}, to solve diffusion-type equations using Particle Strength Exchange (PSE) \cite{Degond:1989a, Eldredge:2002} or Discretization-Corrected PSE (DC-PSE) \cite{Schrader:2010, Bourantas:2016}, for simulations of stochastic models like random walks \cite{Pearson:1905}, and for molecular dynamics simulations of Lennard-Jones dynamics \cite{Lennard-Jones:1931} or deterministic granular flow models \cite{Walther:2009}.
Recently, particle methods have been mathematically defined as an algorithmic class~\cite{Pahlke:2023} and their correctness for distributed computing has been proven~\cite{Pahlke:2024}. This definition opens doors to more theoretical studies of this important class of scientific algorithms, including their computational power.

To study the computational power of particle methods based on their mathematical definition, we interpret them as automata. We then use the tools of automata theory to characterize their computational power. We find that particle methods are Turing complete. 
This can be seen in the simple reduction where an entire universal Turing machine is implemented in the evolve function of the global variable. Such a result is, however, not particularly interesting. More interesting are the two questions: (1) How much can we restrict a particle method such that it remains Turing complete? (2) What is the least restriction of a particle method that causes it to no longer be Turing complete? We address the first question for two sets of restrictions, where one is not a subset of the other. For the second question, we exploit that halting decidability implies loss of Turing powerfulness. Hence, we address this question with a set of restrictions for which prove that the halting problem is decidable for particle methods under these restrictions.

Taken together, the present results provide insight into the computational power of all scientific-computing algorithms that can be subsumed under the class of particle methods. The results are non-trivial and show the limits of Turing powerfulness under sets of restrictions that are directly related to and relevant for practical applications.

\section{Background}
Before proving the present results on Turing powerfulness of particle methods, we introduce the notation used and revisit the basic concepts of particle methods and Turing machines as used in our proofs. This is for the convenience of the reader and serves to make the paper more self-contained.
\subsection{Terminology and Notation}
We introduce the notation and terminology used in the following and define some mathematical concepts. We also briefly recapitulate known concepts for convenience.

\begin{definition} \label{def:def1}\label{def:kleenstar}
	The  \textbf{Kleene star} $A^*$ is the set of all finite tuples of any length with elements from $A$, including the empty tuple $()$. It is defined using the Cartesian product:
	\begin{align}
		A^*:= \bigcup_{j\in \mathbb N_0} A^j \qquad \text{with } A^0\!:=& \{ () \},\quad
		A^1\! := A,\ \
		A^{n+1}\!:= A^n\!\times\! A, \ \ \text{and } n \in \mathbb N_{>0}.
	\end{align}
	
\end{definition}

\begin{notation}
	We use \textbf{bold symbols} for tuples of arbitrary length, e.g.,
	\begin{equation}
		\mathbf p \in P^*.
	\end{equation}
\end{notation}

\begin{notation}
	We use \textbf{regular symbols with subscript indices} for the elements of these tuples, e.g.,
	\begin{equation}
		\mathbf p = (p_1,\ldots ,p_n).
	\end{equation}
\end{notation}

\begin{notation}
	We use \textbf{regular symbols} for tuples of determined length with specific element names, e.g.,
	\begin{equation}
		p = (a,b,c)\in A\times B \times C.
	\end{equation}
\end{notation}

\begin{notation}
	We use the same \textbf{indices} for tuples of determined length and their named elements to identify them, e.g.,
	\begin{equation}
		p_j = (a_j,b_j,c_j).
	\end{equation}
\end{notation}

\begin{definition}
	The \textbf{number of elements of a tuple} $\mathbf{p}=\left(p_1,\ldots ,p_n\right)\in P^*$ is defined as
	\begin{equation}
		\vert \mathbf p \vert := n.
	\end{equation}
\end{definition}

\begin{definition} \label{def:compo}
	The \textbf{composition operator} $*_h$ of a binary function $h: A \times B \rightarrow A$ is recursively defined as:  
	\begin{align}
		&*_h: A\times B^*  \rightarrow A\\[6pt]
		&a *_h () := a\\
		&a *_h (b_1,b_2,\ldots ,b_n) := h(a,b_1) *_h (b_2,\ldots ,b_n).
	\end{align}
\end{definition}

\begin{definition} \label{def:concat}
	The \textbf{concatenation} $\circ : A^* \times A^* \rightarrow A^*$ of tuples $\left(a_1,\ldots ,a_n\right) , \left(b_1,\ldots ,b_m \right)\in A^*$ is defined as:
	\begin{equation}
		\left(a_1,\ldots ,a_n\right) \circ \left(b_1,\ldots ,b_m \right) 	:=  \left(a_1,\ldots ,a_n,b_1,\ldots ,b_m \right).
	\end{equation}
\end{definition}

\begin{definition} \label{def:subtuple}
	We define the \textbf{construction of a subtuple} $\mathbf{b}~\!\in~\!A^*$ of $\mathbf a\in A^*$.
	Let $f: A^* \times \mathbb N \rightarrow \{\top, \bot \} $ ($\top = true$, $\bot = false$) be the condition for an element $a_j$ of the tuple $\mathbf a$ to be in $\mathbf b$.  Then, $\mathbf b=(a_j \in \mathbf a: f(\mathbf a, j))$ defines a subtuple of $\mathbf a$ as:
	\begin{align}
		\begin{split}
			\mathbf{ b}= (a_j \in \mathbf a: f(\mathbf a, j)) := (a_{j_1},\ldots ,a_{j_n})& \\
			\longleftrightarrow \hspace*{10mm} &\mathbf a = (a_1,\ldots , a_{j_1},\ldots ,a_{j_2},\ldots ,a_{j_n},\ldots ,a_m)\\
			\land \quad &\forall k\in \{1,\ldots ,n\}:  \; f(\mathbf a, j_k)=\top\\
			\land \quad  &\forall l\in \{1,\ldots,m\}\backslash \{j_1,\ldots,j_n\}: \; f(\mathbf a, l)=\bot
		\end{split}
	\end{align}
\end{definition}

\begin{definition}\label{def:subresult}
	A \textbf{subresult} ${}_{k}f$ of a function\\ 
	\begin{equation}
		f:A_1\times\ldots\times A_{n} \rightarrow B_1\times\ldots\times B_{m} 
	\end{equation}
	with 
	\begin{equation}
		f\left(a_1,\ldots, a_{n}\right)= \left( b_1,\ldots, b_{m}\right) 
	\end{equation}
	is defined as
	\begin{equation}
		{}_{k}f(a_1,\ldots,a_{n}):= b_k \quad \text{with}\quad k\in \{1,\ldots,m\}.
	\end{equation}
\end{definition}

\subsection{Mathematical Definition of Particle Methods}
The formal definition of a particle method specifies the constituents of the particle method algorithm and their interplay \cite{Pahlke:2023}. The basic data structure is a particle. Particles are collections of properties, such as position, velocity, mass, color, etc. Particles interact in pairs with their neighbors within a certain radius. Then, they individually evolve to change their properties in response to the interactions that have happened. In addition, there is a global variable for readability, which is a collection of non-particle-specific properties, such as the current simulation time, total energy, etc.
Formally, the definition of particle methods consists of three parts: the algorithm, the instance, and the state transition function.  

\subsubsection*{Particle Method Algorithm}\label{sec:defPM:definition} \label{sec:ParticleMethodsDefinition:Algorithm}
The definition of a particle method algorithm encapsulates the structural elements of its implementation in a small set of data structures and functions. 
In detail, the components are:
\begin{definition} This definition is taken from Ref.~\cite{Pahlke:2023}.\\
	A \textbf{particle method algorithm} is a 7-tuple $(P, G, u, f, i, e, \mathring e)$, consisting of the two data structures
	\begin{align}
		\label{eq:defPM:P}
		&P  := A_1 \times A_2 \times \ldots \times A_n 
		&&\text{the particle space,}\\
		\label{eq:defPM:G}
		&G := B_1 \times B_2 \times \ldots \times B_m  
		&&\text{the global variable space,}
	\end{align}
	such that $[G\times P^*]$ 
	is the {\em state space} of the particle method, and the five functions: 
	\begin{align}
		\label{eq:defPM:u}
		&u: [G \times P^*] \times \mathbb N \rightarrow \mathbb N^* 
		&&\text{the neighborhood function,}\\
		\label{eq:defPM:f}
		&f:  G \rightarrow \{ \top,\bot \} 
		&&\text{the stopping condition,}\\
		\label{eq:defPM:i}
		&i:  G \times P \times P \rightarrow P\times P  
		&&\text{the interact  function,}\\
		\label{eq:defPM:e}
		&e:  G \times P\rightarrow G \times P^*  
		&&\text{the evolve function,}  \\
		\label{eq:defPM:eg}
		&\mathring{e} :  G \rightarrow G   
		&&\text{the evolve function of the global variable.}
	\end{align}
\end{definition}
The sets $A_1,\ldots,A_n$ and $B_1,\ldots,B_n$ can be arbitrary, including infinite. For the following computability results, however, we only consider countable sets.

\subsubsection*{Particle Method Instance} \label{sec:defPM:instance}
The particle method instance describes the initial state of a particle method.
\begin{definition} \label{def:PMI} This definition is taken from Ref.~\cite{Pahlke:2023}.\\
	An initial state defines a \textbf{particle method instance} for a given particle method algorithm \sloppy $(P, G, u, f, i, e, \mathring e)$:
	\begin{equation}
		\label{eq:defPMI:gp}[g^1,\mathbf{p}^1] \in [G\times P^*].
	\end{equation}
	The instance consists of an initial value for the global variable $g^1 \in G$ and an initial tuple of particles $\mathbf p^1 \in P^*$.
\end{definition}

\subsubsection*{Particle Method State Transition Function}
The particle method state transition function describes how a particle method proceeds from the instance to the final state by using the particle method algorithm. 
The state transition function consists of a series of state transition steps. This series ends when the stoping function $f$ returns $true$~($\top$). The state transition function is the same for all particle methods, using the functions specified in the corresponding particle method algorithm. 
\begin{definition}This definition is taken from Ref.~\cite{Pahlke:2023}.\\
	The {\em state transition function} $S :  [G\times P^*] \rightarrow [G\times P^*]$ is defined with the Nassi-Shneiderman diagram in Figure~\ref{tab:transitionStd}.
\end{definition}

\newcounter{counter}\stepcounter{counter}
\newcommand{\ct}{\footnotesize{\thecounter\stepcounter{counter}}}
\begin{figure}[!ht]
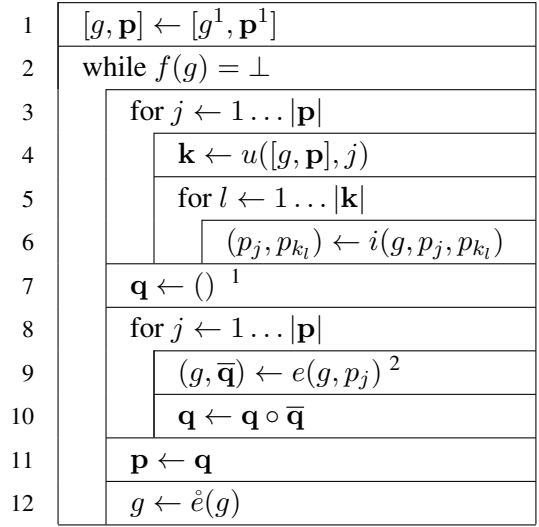

	\centering
	\begin{tabular}{ r| *{4}{l}|}
		\cline{2-5}  
		\ct & \multicolumn{4}{|l|}{$ [g, \mathbf p] \gets [g^1,\mathbf p^1]$  \myh}  \\  
		\cline{2-5} 
		\ct & \multicolumn{4}{|l|}{while $f(g)=\bot$ \myh}  \\ 
		\cline{3-5}
		\ct && \multicolumn{3}{|l|}{for $j \gets 1\ldots \vert \mathbf p \vert$ \myh}  \\ 
		\cline{4-5}
		\ct && \multicolumn{1}{|l}{} & \multicolumn{2}{|l|}{$\mathbf k \gets u([g, \mathbf p], j)$ \myh} \\ 
		\cline{4-5}
		\ct&& \multicolumn{1}{|l}{} & \multicolumn{2}{|l|}{for $l \gets 1\ldots \vert \mathbf k \vert$ \myh} \\ 
		\cline{5-5}
		\ct&& \multicolumn{1}{|l}{} & \multicolumn{1}{|l}{}& \multicolumn{1}{|l|}{$(p_j,p_{k_l}) \gets i(g,p_j, p_{k_l})$ \myh} \\ 
		\cline{3-5}
		\ct&&  \multicolumn{3}{|l|}{$\mathbf{q} \gets () \ $ \footnote{$\mathbf{q}$ is an intermediate result.} \myh}  \\ 
		\cline{3-5}
		\ct&& \multicolumn{3}{|l|}{for $j \gets 1\ldots \vert \mathbf p \vert$ \myh}  \\ 
		\cline{4-5}
		\ct&&  \multicolumn{1}{|l|}{} &\multicolumn{2}{|l|}{$(g,\overline{\mathbf{q}} ) \gets e(g,p_j) \ $\footnote{$\overline{\mathbf{q}}$ is an intermediate result.}\myh}  \\
		\cline{4-5}
		\ct& & \multicolumn{1}{|l|}{} &\multicolumn{2}{|l|}{$\mathbf{q}  \gets \mathbf{q} \circ \overline{\mathbf{q}} $ \myh}  \\
		\cline{3-5}
		\ct& &\multicolumn{3}{|l|}{$\mathbf p \gets \mathbf{q}$\myh}\\
		\cline{3-5}
		\ct&&\multicolumn{3}{|l|}{$g \gets \mathring e(g)$\myh}\\
		\cline{2-5}  
	\end{tabular}
	\caption{Nassi-Shneiderman diagram for the state transition function $S$.} \label{tab:transitionStd}
\end{figure}

The Nassi-Shneiderman diagram corresponds to the following formulas that mathematically define the state transition function $S$. It is divided into sub-functions for better readability.  For each formula, we refer to the lines in the diagram ({\footnotesize 1}-{\footnotesize 12}) corresponding to it. All interact sub-functions return an altered particle tuple $\mathbf{p}$. 
The first interact sub-function $\intI$ calculates one pairwise particle interaction and formalizes line {\footnotesize 6}:
\begin{multline}
	\label{eq:defPM:si1}
	\intI ([g,\mathbf p], j,k) :=(p_1,\ldots ,p_{j-1},\overline p_j, p_{j+1},\ldots ,p_{k-1},\overline p_k p_{k+1},\ldots ,p_{\vert\mathbf p\vert}),  \\
	\hspace{13mm} \text{with }\quad   (p_1,\ldots ,p_{\vert \mathbf p\vert}) = \mathbf p , \quad\left(\overline p_j, \overline p_k\right) := i\left( g,p_j,p_k\right).
\end{multline}
The second interact sub-function $\intII$ computes the interaction of one particle with all of its neighbors and formalizes lines {\footnotesize 4} to {\footnotesize 6}:
\begin{equation}
	\label{eq:defPM:si2}
	\intII ([g,\mathbf p], j) := \mathbf p \; *_{\intI_{(g,j)}} \; u([g,\mathbf p], j)\
	\hspace{13mm}\text{ with } \quad \intI_{(g,j)}(\mathbf p,k):=\intI ([g,\mathbf p], j,k), 
\end{equation}
The third interact sub-function $\intIII$ computes the interaction of all particles with all of their respective neighbors and formalizes lines {\footnotesize 3} to {\footnotesize 6}:
\begin{equation}
	\label{eq:defPM:si3}
	\intIII ([g,\mathbf p]):=\mathbf p \; *_{\intII_g} (1,\ldots ,\vert \mathbf p\vert) \
	\hspace{13mm} \text{with } \quad \intII_g (\mathbf p, j):=\intII ([g,\mathbf p], j).
\end{equation}
The first evolution sub-function $\evoI$ computes the evolution of one particle and stores the result in an intermediate particle tuple $\mathbf{q}$ and in the global variable. It formalizes lines {\footnotesize 9} to {\footnotesize 10}:
\begin{equation}
	\label{eq:defPM:se1}
	\evoI \big(  [g,\mathbf p], \mathbf q, j \big):= \big[ \overline g, \quad \mathbf q \circ \overline{\mathbf q}   \big]   \
	\hspace{13mm} \text{with } \quad \left(\overline g, \overline{\mathbf q}\right) := e(g,p_j).
\end{equation}
The second evolution sub-function $\evoII$ computes the evolution of all particles and returns the result in a new particle tuple $\mathbf{q}$ and in the global variable. It formalizes lines {\footnotesize 7} to {\footnotesize 10}:
\begin{equation}
	\label{eq:defPM:se2}
	\evoII \big(  [g,\mathbf p]\big) := \big[ g, () \big] \; *_{\evoI_{\mathbf p}} \; (1,\ldots ,\vert \mathbf p\vert)
	\
	\hspace{13mm} \text{with} \quad \evoI_{\mathbf p} \big( [g,\mathbf q], j \big)  := \evoI \big( [g,\mathbf p], \mathbf q, j \big).
\end{equation}
The state transition step $s$ composes all sub-functions  and formalizes lines {\footnotesize 3} to {\footnotesize 12}:
\begin{equation}
	\label{eq:defPM:ss}
	s \left(  [g, \mathbf p] \right):=
	\big[ \mathring{e}( \overline{ g} ),\; \overline{\mathbf p}   \big]
	\qquad \text{with} 
	\quad  \left[\overline g, \overline{\mathbf p}\right] := \evoII \big(  [g,\; \intIII ([g,\mathbf p])]\big).
\end{equation}
Finally, the state transition function $S$ advances the algorithm to the final state and formalizes lines {\footnotesize 1} to {\footnotesize 12}:
\begin{multline}\label{eq:stateTransitionFunctionDefinition}
	S([ g^1, \mathbf p^1]) = [ g^T, \mathbf p^T]  
	\qquad\longleftrightarrow \\
	f(g^T)=\top \quad\
	\land \quad \forall t\in \{2,\ldots ,T\}:\ [g^{t},\mathbf{p}^{t}]=s\left([g^{t-1},\mathbf{p}^{t-1}]\right)\ \land \ f(g^{t-1})=\bot.
\end{multline}

\subsection{Turing Machines}\label{sec:TuringMachineByKozen}
Turing machines \cite{Turing1936} are the most general and powerful automata known so far.
There exist many equivalently powerful formulations of Turing machines. We follow the formulation of Kozen\footnotemark~\cite{kozen2012automata}.
\footnotetext{A Turing machine consists of a semi-finite tape, a read-write head, and a state. The tape consists of cells. Each cell stores one symbol from the tape alphabet $\Gamma$.
The cell at the beginning of the tape contains the end marker $\vdash$.
Initially, a finite string of input symbols from the input alphabet $ \Sigma \subseteq \Gamma$ extends to the right of the end marker. The tape's infinite rest is filled with blank symbols $\blank$.
The read-write head points at exactly one cell at a time. Initially, it points at the first cell, containing the end marker.
Depending on the symbol in the cell currently pointed to, and on the current state of the Turing machine, the head changes the state of the Turing machine, overwrites the symbol in the cell, and moves right ($+1$) or left ($-1$) by one cell, as determined by the transition function $\delta$.}
Formally, the definition of Turing machines used here is:

\begin{definition} This definition is taken from the classic book by Dexter Kozen~\cite{kozen2012automata}.\\
	\sloppy
	A  \textbf{deterministic one-tape Turing machine} is a tuple 
	\begin{equation}
		\left(Q,\Sigma, \Gamma, \vdash, \blank, \delta, \textbf{start}, \textbf{accept}, \textbf{reject}\right)
	\end{equation} such that:
	\begin{align} 
		Q &\text{ is a finite set of states, } \\
		\Sigma &\text{ is a finite input alphabet, } \\
		\Gamma &\text{ is a finite tape alphabet with } \Sigma \subseteq \Gamma,\\
		\vdash  &\text{ is the end marker with } \vdash  \in \Gamma \backslash \Sigma,\\
		\blank & \text{ is the blank symbol with } \blank  \in \Gamma \backslash \Sigma,\\
		\delta&\text{ is the transition function with }\
		\delta : Q\times \Gamma \rightarrow Q\times \Gamma\times \{-1,+1\}, \\
		\textbf{start} &\text{ is the start state}, \\
		\textbf{accept} &\text{ is the accept state}, \\
		\textbf{reject} &\text{ is the reject state with }\textbf{reject} \neq \textbf{accept}.
	\end{align}
	The head is not allowed to overwrite the end marker or leave the tape:
	\begin{equation}\label{eq:defTM:delta:CellOneRule}
		\forall q\in Q \,\,\exists \, \overline q \in Q :  \delta(q,\vdash) = (\overline q, \vdash, 1).
	\end{equation}
	Additionally, a Turing machine must stay in the accept or reject state once one of those states has been entered:
	\begin{align}
		\forall b\in \Gamma \,\, \exists \, c,\overline c \in \Gamma \land d,\overline d \in \{-1,+1\}:
		\delta(\textbf{accept},b) &= (\textbf{accept},c,d)\\
		\land\,\,  \delta(\textbf{reject},b) &= (\textbf{reject},\overline c,\overline d).
	\end{align}
\end{definition}

\begin{definition} This definition is taken from Ref.~\cite{kozen2012automata}.\\
	A  \textbf{configuration} of a Turing machine is a tuple:
	\begin{equation}
		(q,\vdash \mathbf x \blank^{\omega},n) \in C:=Q \times \{\vdash \mathbf x \blank^{\omega}: \mathbf x\in\Gamma^*\}\times \mathbb N,
	\end{equation}
	where
	\begin{align} 
		q &\text{ is the state,} \\
		\vdash \mathbf x \blank^{\omega} &\text{ is the semi-finite string\footnotemark \ on the tape, and} \\
		n &\text{ is the head's position on the tape.}
	\end{align}
	A configuration contains all information about a Turing machine's current state.
	\footnotetext{The semi-finite string $\vdash \mathbf x \blank^{\omega}$ contains the end marker $\vdash$, a finite string $\mathbf x\in \Gamma^*$, and the semi-finite string $\blank^{\omega}$ where $\omega$ is the smallest ordinal number.}

	The  \textbf{start configuration} of a Turing machine  is:
	\begin{equation}\label{eq:TuringMachineByKozen:startConfiguration}
		(\textbf{start}, \mathbf x^1 \blank^{\omega},1) \text{~~with~~} \mathbf x^1 =\ \vdash\mathbf{y} \text{~~and~~}\mathbf{y}\in \Sigma^* ,
	\end{equation}
	where $\mathbf y$ is the finite input string of the Turing machine and $1$ means that the head is in the first cell containing $\vdash$.\end{definition}

\begin{definition} This definition is taken from Ref.~\cite{kozen2012automata}.\\
	Be $x_n$ the $n$-th symbol of the string $\mathbf x$, where $x_1$ is the left-most symbol, and be $s^n_b(\mathbf x)$ the string where the $n$-th symbol of $\mathbf x$ is replaced by $b$. Then, the  \textbf{next-configuration relation $\underset{M}{\overset{1}{\longrightarrow}}$} of a Turing machine is defined by:
	\begin{equation}
		\underset{M}{\overset{1}{\longrightarrow}} \,\subseteq C\times C
	\end{equation}
	\begin{equation}
		\Big((q,\mathbf x,n), ( q',\mathbf x', n')\Big)\in\, \underset{M}{\overset{1}{\longrightarrow}}\,\,
		\longleftrightarrow \ 
		( q',\mathbf x', n')=( \overline q, s^n_b(\mathbf x), n+d)  \text{~~~for~~}  \delta(q,x_n)=( \overline q, b, d).
	\end{equation}
	Since we consider deterministic Turing machines, $\underset{M}{\overset{1}{\longrightarrow}}$ is a function.
\end{definition}

\section{Turing Powerfulness of Particle Methods}
Particle methods in general are trivially Turing complete, as one can implement a universal Turing machine in the global variable and the evolve function of the global variable. Hence, a more interesting question is under what set of restrictions a particle method is still Turing powerful. Since sets of restrictions do not possess a total order, we provide proofs of Turing powerfulness for two sets of restrictions, where one is not a subset of the other.
We use the term \textit{Turing powerfulness} to indicate that we are not proving Turing completeness. Indeed, we prove that even restricted particle methods are at least as powerful as Turing machines, but not necessarily vice versa.

\subsection{First Set of Restrictions}\label{sec:first}
\begin{theorem}\label{thm:TMsimple}
	Any particle method that fulfills all of the following restrictions is Turing powerful:
	({\em Note that we use subresult notation as defined in Def.~\ref{def:subresult}.})
	\begin{enumerate}
		\item It has an empty neighborhood:
		\begin{equation}\label{eq:TMsimple:condition:u}
			u\left([g,\mathbf{p}],j\right)=().
		\end{equation}
		
		\item The interact function is the identity:
		\begin{equation}\label{eq:TMsimple:condition:i}
			i\left(g,p_j,p_k\right)=\left(p_j,p_k\right).
		\end{equation}
		
		\item The evolve function is order-independent w.r.t.~$g$:
		\begin{equation}\label{eq:TMsimple:condition:e:g}
			{}_{1}e\left({}_1 e\left(g,p'\right),p''\right)={}_{1}e\left({}_1 e\left(g,p''\right),p'\right).
		\end{equation}
		
		\item The evolve function is independent of previous evolutions of $p$:
		\begin{equation}\label{eq:TMsimple:condition:e:p}
			{}_{2}e\left(g,p'\right)={}_{2}e\left({}_1 e\left(g,p''\right),p'\right)'
		\end{equation}
		
		\item The size of the global variable $\varsigma_g$, the size of each particle $\varsigma_p$, the time complexity of the stop function $\tau_{f}$, of the evolve function $\tau_e$, and of the evolve function of the global variable $\tau_{\overset{\circ }{e}}$ are all in ${O}(\log(|\mathbf{p}^t|))$: 
		\begin{equation}\label{eq:TMsimple:condition:complexity}
			\varsigma_{p_j^t} ,
			\tau_{f} ,
			\tau_{e}, 
			\tau_{\overset{\circ }{e}}
			\in {O}\left(\log(|\mathbf{p}^t|)\right) .
		\end{equation}
	\end{enumerate}
\end{theorem}

\begin{proof}
	To prove the above statement, we use the overbar notation 
	\begin{equation}
		(\overline{\mathfrak q},\overline{\mathfrak z},\overline{\mathfrak d}):=\delta (\mathfrak q,z)
	\end{equation}
	for the results of the transition function.
	Without loss of generality, we assume a strict total order $<$ on the set of states $Q$ of the Turing machine, where
	\begin{equation}
		\forall q \in Q\backslash \{\textbf{start}\}: \textbf{start} < q.
	\end{equation}
	We construct a particle method that fulfills the constraints and emulates an arbitrary Turing machine as defined in section \ref{sec:TuringMachineByKozen}.
	The constructed particle method for a Turing machine that fulfills the constraints in Eqs.~(\ref{eq:TMsimple:condition:u})--(\ref{eq:TMsimple:condition:complexity}), as proven in Appendix \ref{app:A}, is:
	\begin{align}
		\label{eq:TMsimple:p}
		p:= & \left(\mathfrak k,\mathfrak z\right) \ \ \text{ for } p\in P:=\mathbb{N}_{>0}\times \Gamma
		\\
		\label{eq:TMsimple:g}
		g:= & \left(\mathfrak q,\Delta\mathfrak q,\mathfrak d,\mathfrak m,\mathfrak M\right) \ \ \text{ for } p\in P:=Q^2 \times \{-1,+1\}\times {\mathbb{N}_{>0}}^2
		\\
		\label{eq:TMsimple:u}
		u\left([ g,\mathbf{p}] ,j\right) := & \left( \right)
		\\
		\label{eq:TMsimple:f}
		f( g) := & \left(\mathfrak q =\textbf{reject} \ \lor \mathfrak q =\textbf{accept} \}\right)
		\\
		\label{eq:TMsimple:i}
		i( g,p_{j} ,p_{k}) := & \left(p_{j} ,p_{k}\right)
		\\
		\label{eq:TMsimple:e}
		e( g,p_{j}) :=&\begin{cases}
			\left(\left(\mathfrak q,\max( \Delta\mathfrak q,\overline{\mathfrak q}) ,\max(\mathfrak d,\overline{\mathfrak d}) ,\mathfrak m,\mathfrak M\right) ,\left(( \mathfrak k_{j} ,\overline{\mathfrak z})\right)\right) &\text{if } \mathfrak k_{j} =\mathfrak m\land \mathfrak m+\overline{\mathfrak d} \leq \mathfrak M
			\\
			\Big(\left(\mathfrak q,\max( \Delta\mathfrak q,\overline{\mathfrak q}) ,\max( \mathfrak d,\overline{\mathfrak d}) ,\mathfrak m,\mathfrak M\right) , & 
			\\
			\ \left(( \mathfrak k_{j} ,\overline{\mathfrak z}) ,( \mathfrak k_{j} +1,_{-})\right)\Big) & \text{if } \mathfrak k_{j} =\mathfrak m\land \mathfrak m+\overline{\mathfrak d}  >\mathfrak M\\
			\left( g,( p_{j})\right) & \text{else}
		\end{cases}
		\\
		\label{eq:TMsimple:eg}
		\overset{\circ }{e}( g) := &\begin{cases} \left(\Delta\mathfrak q,\textbf{start},-1,\mathfrak m+\mathfrak d, \mathfrak M+1\right)& \text{if } \mathfrak m+\mathfrak d > \mathfrak M \\
			\left(\Delta\mathfrak q,\textbf{start},-1,\max(1,\mathfrak m+\mathfrak d), \mathfrak M\right)	&\text{else.}
		\end{cases}
	\end{align}
	
	In this particle method, each particle represents one cell on the tape. The particle properties store the cell index $\mathfrak{k}$ and the symbol $\mathfrak{z}$. The global variable stores the Turing machine's read-write head positions and contains the properties for orchestrating the creation of new particles (tape cells). Therefore, it contains the state of the Turing machine $\mathfrak q$, an accumulator for the state $\Delta\mathfrak q$, the movement direction of the head $\mathfrak d$, the position (index) of the head $\mathfrak{m}$, and the total number of particles $\mathfrak{M}$. The accumulator $\Delta\mathfrak q$ is necessary for the order-independence of the evolve function. 
	The neighborhood is empty. Hence, there are no interactions. The evolve function only differs from the identity for the particle to which the head currently points ($p_\mathfrak{m}$). Then, the transition function $\delta$ is evaluated, and the result is stored in the global variable and in the particle, respectively. 
	The state $\mathfrak{q}$ is not updated, to make the evolve function independent of previous evolutions. The $\max$ function for overwriting the accumulator $\Delta\mathfrak q$ and the direction $\mathfrak{d}$ renders the evolve function order-independent. 
	If in the new state the head points to a non-existant particle (i.e., empty cell) $\mathfrak m+\overline{\mathfrak d}  >\mathfrak M$, this particle is created by the evolve function with the blank symbol $\blank$ as its property.
	The evolve function of the global variable  $\overset{\circ }{e}$ changes the position of the head $\mathfrak{m}$ according to the head movement $\mathfrak{d}$, adds $1$ to the number of particles if a new particle was created, and resets the accumulator $\Delta\mathfrak q$ and the direction $\mathfrak{d}$ to the smallest values, i.e., $\textbf{start}$ and $-1$.
	
\end{proof}

\subsection{Second Set of Restrictions}\label{sec:second}
\begin{theorem}\label{thm:TMcomplex}
	Any particle method that fulfills all of the following restrictions is Turing powerful:\\
	({\em Note that we use subresult notation as defined in Def.~\ref{def:subresult}.})
	\begin{enumerate}
		
		\item $P$ and $G$ are finite:
		\begin{equation}\label{eq:TMcomplex:condition:PG}
			|P|<\infty, \quad |G|<\infty \, .
		\end{equation}
		
		\item The interact function is a pull interaction (i.e., only the first particle $p_j$ is changed):
		\begin{equation}\label{eq:TMcomplex:condition:i:pull}
			i\left(g,p_j,p_k\right)=\left(\overline p_j,p_k\right).
		\end{equation}
		
		\item The interact function is independent of previous interactions:
		\begin{equation}\label{eq:TMcomplex:condition:i:Previous}
			_{1} i_{g} (p_{j} ,{}_{1} i_{g} (p_{k} ,p_{k'} ))={}_{1} i_{g} (p_{j} ,p_{k} ).
		\end{equation}
		
		\item The interact function is order-independent:
		\begin{equation}\label{eq:TMcomplex:condition:i:Order}
			_{1} i_{g} ({}_{1} i_{g} (p_{j} ,p_{k} ),p_{k'} )={}_{1} i_{g} ({}_{1} i_{g} (p_{j} ,p_{k'} ),p_{k} ).
		\end{equation}
		
		\item The evolve function is order-independent w.r.t.~$g$:
		\begin{equation}\label{eq:TMcomplex:condition:e:order}
			{}_{1}e\left({}_1 e\left(g,p'\right),p''\right)={}_{1}e\left({}_1 e\left(g,p''\right),p'\right).
		\end{equation}
		
		\item The evolve function is independent of previous evolutions of $p$:
		\begin{equation}\label{eq:TMcomplex:condition:e:previous}
			{}_{2}e\left(g,p'\right)={}_{2}e\left({}_1 e\left(g,p''\right),p'\right).
		\end{equation}
		
		\item The neighborhood function is independent of $g$ and of the values of $p_j\in\mathbf{p}$:
		\begin{equation} \label{eq:TMcomplex:condition:u:valueindependence}
			u([g,\mathbf p], j)= (k\in \{1,\ldots ,|\mathbf{p}|\}:\Omega(j,k)=\top).
		\end{equation}
		
		\item The neighborhood function is independent of previous interactions:
		\begin{equation} \label{eq:TMcomplex:condition:u:previous}
			u([g,\mathbf p], j)= u([g,\mathbf p *_{\iota^I_{(g,k')}} (k'')], j).
		\end{equation}

		\item $e$, $i$, $f$, and $\overset{\circ }{e}$ have a time complexity that is bounded by a constant:
		\begin{equation}\label{eq:TMcomplex:condition:functions:timecomplexity}
			\tau_e,\tau_i,\tau_u,\tau_f,\tau_{\overset{_\circ}{e}}\in O(1).
		\end{equation}
		
		\item  $e$, $i$, $u$, $f$, and $\overset{\circ }{e}$ have a space complexity that is bounded by a constant:
		\begin{equation}\label{eq:TMcomplex:condition:functions:spacecomplexity}
			\varsigma_e,\varsigma_i,\varsigma_u,\varsigma_f,\varsigma_{\overset{_\circ}{e}}\in O(1).
		\end{equation}
	\end{enumerate}	
\end{theorem}

\begin{proof}
	We again use the overbar notation 
	\begin{equation}
		(\overline{\mathfrak{q}}_{j} ,\overline{\mathfrak{z}}_{j} ,\overline{\mathfrak{d}}_{j} ):=\delta (\mathfrak{q} ,\mathfrak{z}_{j} )
	\end{equation}
	and assume w.l.o.g.~a strict total order $<$ on the set of states $Q$ of the Turing machine, where
	\begin{equation}
		\forall q \in Q\backslash \{\textbf{start}\}: \textbf{start} < q.
	\end{equation}
	We construct a particle method that fulfills the constraints and emulates an arbitrary Turing machine as defined in section \ref{sec:TuringMachineByKozen}.
	The constructed particle method for a Turing machine that fulfills the constraints in Eqs.~(\ref{eq:TMcomplex:condition:PG}) to (\ref{eq:TMcomplex:condition:functions:spacecomplexity}), as proven in Appendix \ref{app:B}, is:
	\begin{equation}
		p:=(\mathfrak{z,h} ,\Delta \mathfrak{h} ,\mathfrak{o} ,\Delta \mathfrak{o} ,\mathfrak{a} )\ \ \text{ for } p\in P:=\underbrace{\Gamma }_{\mathfrak{z} \ \in } \times \underbrace{\{-1,0,1\}}_{\mathfrak{h} \ \in } \times \underbrace{\{0,1\}}_{\Delta \mathfrak{h} \ \in } \times {\underbrace{\{-1,1\}}_{\mathfrak{o} ,\ \Delta \mathfrak{o} \ \in }}^{2} \times \underbrace{\{0,1\}}_{\mathfrak{a} \ \in }
	\end{equation}
	
	\begin{equation}
		g:=(\mathfrak{q} ,\Delta \mathfrak{q} )\ \text{ for } \ g\in G:=Q\times Q
	\end{equation}
	
	\begin{equation}
		u([g,\mathbf{p} ],j):=(k\in (1,...,|\mathbf{p} |):k=j-1\lor k=j+1)
	\end{equation}
	
	\begin{equation}
		f(g):=\left(\mathfrak{q} =\textbf{reject} \ \lor \mathfrak{q} =\textbf{accept} \}\right)
	\end{equation}
	
	\begin{align}
		& i(g,p_{j} ,p_{k} ):=\\
		& \begin{cases}
			\biggl(\Bigl(\mathfrak{z}_{j}\mathfrak{,h}_{j} ,\underbrace{1}_{\Delta \mathfrak{h}_{j}} ,\mathfrak{o}_{j} ,\underbrace{\max (\Delta \mathfrak{o}_{j} ,\overline{\mathfrak{d}}_{k} )}_{\Delta \mathfrak{o}_{j}} ,\mathfrak{a}_{j}\Bigr) ,\ p_{k}\biggr) & \begin{array}{ l c }
				\text{if} & \mathfrak{h}_{k} =1\land ((\mathfrak{h}_{j} =-1\mathfrak{\land }\overline{\mathfrak{d}}_{k} \neq \mathfrak{o}_{k} )\\
				& \lor (\mathfrak{h}_{j} =0\mathfrak{\land }\overline{\mathfrak{d}}_{k} =\mathfrak{o}_{k} ))
			\end{array}\\
			\biggl(\Bigl(\mathfrak{z}_{j} ,\mathfrak{h}_{j} ,\Delta \mathfrak{h}_{j} ,\mathfrak{o}_{j} ,\Delta \mathfrak{o}_{j} ,\underbrace{1}_{\mathfrak{a}_{j}}\Bigr) ,\ p_{k}\biggr) & \begin{array}{ l c }
				\text{if} & \mathfrak{h}_{j} =1\land ((\mathfrak{h}_{k} =-1\mathfrak{\land }\overline{\mathfrak{d}}_{j} \neq \mathfrak{o}_{j} )\\
				& \lor (\mathfrak{h}_{k} =0\mathfrak{\land }\overline{\mathfrak{d}}_{j} =\mathfrak{o}_{j} ))
			\end{array}\\
			(p_{j} ,\ p_{k} ) & \text{else}
		\end{cases}
	\end{align}
	
	\begin{align}
		& e(g,p_{j} ):=\\
		& \begin{cases}
			\Biggl( g,\biggl(\Bigl(\mathfrak{z_{j} ,}\underbrace{1}_{\mathfrak{h}_{j}} ,\underbrace{0}_{\Delta \mathfrak{h}_{j}} ,\underbrace{\Delta \mathfrak{o}_{j}}_{\mathfrak{o}_{j}} ,\underbrace{-1}_{\Delta \mathfrak{o}_{j}} ,\underbrace{0}_{\mathfrak{a}_{j}}\Bigr)\biggr)\Biggr) & \text{if } \Delta \mathfrak{h}_{j} =1\\
			\Bigl(g,\Bigl(\Bigl(\mathfrak{z}_{j} ,\underbrace{0}_{\mathfrak{h}_{j}} ,\underbrace{0}_{\Delta \mathfrak{h}_{j}} ,\underbrace{-1}_{\mathfrak{o}_{j}} ,\underbrace{-1}_{\Delta \mathfrak{o}_{j}} ,\underbrace{0}_{\mathfrak{a}_{j}}\Bigr) \Bigr)\Bigr) & \text{if } \Delta\mathfrak{h}_{j} =0\land \mathfrak{h}_{j} =-1 \\
			\Biggl(\biggl(\mathfrak{q} ,\underbrace{\max (\Delta \mathfrak{q} ,\overline{\mathfrak{q}}_{j} )}_{\Delta \mathfrak{q}}\biggr) ,\biggl(\Bigl(\underbrace{\overline{\mathfrak{z}}_{j}}_{\mathfrak{z}_{j}} ,\underbrace{-1}_{\mathfrak{h}_{j}} ,\underbrace{0}_{\Delta \mathfrak{h}_{j}} ,\underbrace{-1}_{\mathfrak{o}_{j}} ,\underbrace{-1}_{\Delta \mathfrak{o}_{j}} ,\underbrace{0}_{\mathfrak{a}_{j}}\Bigr)\biggr)\Biggr) & \text{if }\Delta\mathfrak{h}_{j} =0\land\mathfrak{h}_{j} =1\land \mathfrak{a}_{j} =1\\
			\Biggl(\biggl(\mathfrak{q} ,\underbrace{\max (\Delta \mathfrak{q} ,\overline{\mathfrak{q}}_{j} )}_{\Delta \mathfrak{q}}\biggr) ,\biggl(\Bigl(\underbrace{\overline{\mathfrak{z}}_{j}}_{\mathfrak{z}_{j}} ,\underbrace{-1}_{\mathfrak{h}_{j}} ,\underbrace{0}_{\Delta \mathfrak{h}_{j}} ,\underbrace{-1}_{\mathfrak{o}_{j}} ,\underbrace{-1}_{\Delta \mathfrak{o}_{j}} ,\underbrace{0}_{\mathfrak{a}_{j}}\Bigr) , & \text{if }\Delta\mathfrak{h}_{j} =0\land\mathfrak{h}_{j} =1\land \mathfrak{a}_{j} =0\\
			\Bigl(\underbrace{\blank}_{\mathfrak{z}}\mathfrak{,}\underbrace{1}_{\mathfrak{h}} ,\underbrace{0}_{\Delta \mathfrak{h}} ,\underbrace{\overline{\mathfrak{d}}_{j}}_{\mathfrak{o}} ,\underbrace{0}_{\Delta \mathfrak{o}} ,\underbrace{0}_{\mathfrak{a}}\Bigr)\biggr)\Biggr) & \\
			\Bigl(g,(p_j)\Bigr) & \text{else}
		\end{cases}
	\end{align}
	
	\begin{equation}
		\overset{\circ }{e} (g):=\Big(\underbrace{\Delta \mathfrak{q}}_{\mathfrak{q}} ,\underbrace{\textbf{start}}_{\Delta \mathfrak{q}}\Big)
	\end{equation}
	
	In this particle method, each particle represents one cell on the tape. All functions except the neighborhood function are bounded by a constant time complexity. Therefore, it is not possible to perform any computations involving particle indices, as their number grows with the number of particles. Indices can be avoided, however, by instead using the previous location of the head together with the information on where it moved from there.
	Both are stored in the particle property $\mathfrak{h}$. The particle with $\mathfrak{h}=1$ marks the current position of the head. The particle with $\mathfrak{h}=-1$ is the position of the head in the previous iteration. On all other particles is $\mathfrak{h}=0$. The property $\Delta\mathfrak{h}$ stores changes to the head position, in order to render the interact function $i$ independent of previous interactions. This information is only stored on one of the particles and not in the global variable, which cannot store an index due to the size constraint.
	The direction of the last move of the head is stored in $\mathfrak{o}$ on the particle where $\mathfrak{h}=-1$. If $\mathfrak{o}=-1$, the head moved left from there. If $\mathfrak{o}=1$, it moved to the right. The property $\Delta\mathfrak{o}$ stores the direction of the next move of the head, only on the particle where the head will be next. Again, this helps render the interact function independent of previous interactions. The global variable stores the state of the Turing machine. 
	
	The interact function $i$ implements the exchange of information about the head position. Hence, $i$ is not the identity for interactions between the particle with $\mathfrak{h}=1$ and the particle to where the head goes. Collectively, this information about the head movement allows determining whether the head goes back to where it came from or moves on, without needing to know the cell indices. Also, new particles can thus be created whenever the head moves beyond the current particle tuple. In such a case, the flag $\mathfrak{a}$ stays $0$ for the particle with $\mathfrak{h}=1$, which is used in the evolve function to create a new particle. The $\max$ reduction when determining $\Delta\mathfrak{o}$ ensures that the interact function is order-independent.
	
	Besides creating new particles, the evolve function $e$ is also responsible for computing the properties of the next state. This includes, that the evolve function changes the global variable according to the transition function of the Turing machine $\delta$. Hence, changing $\Delta\mathfrak{q}$ to the new state of the Turing machine $\overline{\mathfrak{q}}$ and the tape symbol of the particle with the head ($\mathfrak{h}=1$) form $\mathfrak{z}$ to $\overline{\mathfrak{z}}$. 
	In addition, this also includes making the particle that contains the head next (i.e., $\Delta\mathfrak{h}=1$) as the current head position $\mathfrak{h}=1$ and marking the particle that contained the head in the previous step and did not get it back (i.e., $\Delta\mathfrak{h}=0$, $\mathfrak{h}=-1$) as a standard particle with no head information $\mathfrak{h}=0$. Further, the particle that currently has the head and was able to hand it to a neighboring particle (i.e., $\Delta\mathfrak{h}=0$, $\mathfrak{h}=1$, $\mathfrak{a}=1$) is made to the particle that had previously the location of the head $\mathfrak{h}=-1$. Finally, the particle that currently has the head and was not able to hand it to another particle (i.e., $\Delta\mathfrak{h}=0$, $\mathfrak{h}=1$, $\mathfrak{a}=0$) is marked as a particle that had the head $\mathfrak{h}=-1$, and a new particle is created that now has the head $\mathfrak{h}=1$. All other particles remain unchanged.
	The separate accumulator for the state change, $\Delta\mathfrak{q}$, renders the evolve function independent of previous evolutions of $p$, and the $\max$ reduction guarantees order-independence w.r.t.~$g$.
	
	Finally, the evolve function of the global variable, $\overset{\circ }{e}$, sets the state of the Turing machine $\mathfrak{q}$ to $\Delta\mathfrak{q}$ and resets the accumulator $\Delta\mathfrak{q}$ to the minimal value $\textbf{start}$.
	
\end{proof}

\section{Halting Decidability of Particle Methods}
The halting problem is the question of deciding whether a given algorithm halts for a given input. This question is generally undecidable for Turing machines \cite{Turing1936}. Since particle methods are Turing powerful even under non-overlapping restrictions (Theorems~\ref{thm:TMsimple} and \ref{thm:TMcomplex}), the halting problem is, in general, undecidable for a particle method, even if non-trivially restricted. We therefore next answer the question: ``What is a sufficient set of restrictions under which the halting problem becomes decidable for a particle method?''

\begin{theorem}
	The halting problem is decidable for particle methods that fulfill all of the following restrictions:
	\begin{enumerate}
		\item The number of particle properties and of global-variable properties are finite, and all properties are from finite sets:
		\begin{equation}\label{eq:HaltingProblem1:conditionFiniteProperties}
			|P|<\infty \, , \quad |G|<\infty.
		\end{equation}
		\item 
		\begin{equation}\label{eq:HaltingProblem1:conditionComputable}
			e, i, u, f, \text{and } \overset{\circ }{e} \text{ are computable functions.}	
		\end{equation}
		\item The evolve function is not allowed to produces new particles:
		\begin{equation}\label{eq:HaltingProblem1:conditionNoParticleProduction}
			|{}_{2}e(g, p_j)|\leq 1.
		\end{equation}
	\end{enumerate}
\end{theorem}

{\em Note:} A function over a finite discrete space can be represented by a table of input-output pairs. However, it is conceivable that one of the functions implements a Turing machine. It is then not clear if the function terminates. Hence, restriction (2) does not follow from restriction (1), but is a necessary prerequisite to being able to write an input-output table for the function.

\begin{proof}
	For all of the following steps, the conditions in Eq.~(\ref{eq:HaltingProblem1:conditionFiniteProperties}) are essential. We show halting decidability by showing that under the above restrictions, the state transition function is computable with a finite number of possible outcomes.
	
	Since the interact function $i$ is computable (Eq.~(\ref{eq:HaltingProblem1:conditionComputable})) and the initial particle tuple $\mathbf p^1$ is finite (Eq.~(\ref{eq:defPMI:gp}), Def.~\ref{def:kleenstar}), the first interact sub-function $\intI$ (Eq.~(\ref{eq:defPM:si1})) is computable.
	
	The second interact sub-function $\intII$ (Eq.~(\ref{eq:defPM:si2})) is also computable, because the first interact sub-function is computable, the neighborhood function is computable (Eq.~(\ref{eq:HaltingProblem1:conditionComputable})) and the neighborhood size is finite (Eq.~(\ref{eq:defPM:u}), Def.~\ref{def:kleenstar}).
	
	The third interact sub-function $\intIII$ (Eq.~(\ref{eq:defPM:si3})) is computable, because the second interact sub-function $\intII$ is computable and the initial particle tuple $\mathbf p^1$ is finite (Eq.~(\ref{eq:defPMI:gp})).
	
	The first evolve sub-function $\evoI$ (Eq.~(\ref{eq:defPM:se1})) is computable, because the evolve function is computable.
	
	The second evolve sub-function $\evoII$ (Eq.~(\ref{eq:defPM:se2})) is computable, because the first evolve sub-function $\evoI$ is computable and the initial particle tuple $\mathbf p^1$ is finite (Eq.~(\ref{eq:defPMI:gp})).
	
	The step function $s$ (Eq.~(\ref{eq:defPM:ss})) is computable, because the third interact sub-function $\intIII$, the second evolve function $\evoII$, and the evolve function of the global variable $\overset{\circ }{e}$ are all computable (Eq.~(\ref{eq:HaltingProblem1:conditionComputable})).
	
	To decide if a particle method state is a final state, the stop function $f$ is needed. It is also computable (Eq.~(\ref{eq:HaltingProblem1:conditionComputable})) and, hence, decidable.
	
	The initial particle tuple $\mathbf p^1$ is finite (Eq.~(\ref{eq:defPMI:gp})) with a size of $m^1:=|\mathbf{p}^1|$, and the evolve function $e$ cannot produce particles (Eq.~(\ref{eq:HaltingProblem1:conditionNoParticleProduction})). Since this is the only place where particles could potentially be produced, the number of particles in subsequent steps of the algorithm is always smaller than or equal to that of the previous step (i.e., $m^{t+1}\leq m^t$). From this, we can define the set of all states that can be reached by the algorithm from an instance with $m^1=|\mathbf{p}^1|$:
	\begin{equation}
		[G\times P^{*{<m}}] \quad \text{for }\ P^{*{<m}} := \bigcup_{j=0}^{m} P^j.
	\end{equation}
	From $m^1< \infty$ (Eq.~(\ref{eq:defPMI:gp})) and $|P|$, $|G|$ finite (Eq.~(\ref{eq:HaltingProblem1:conditionFiniteProperties})), it follows that 
	\begin{equation}
		M:=\Big|[G\times P^{*<m}]\Big|<\infty.
	\end{equation}
	Hence, there are only finitely many reachable states for any given instance. Since the particle method is assumed to be deterministic, we therefore know after a maximum of $M$ state transition steps whether states are repeating or the particle method stops, i.e., it does not halt or it halts. 
	This means that we can decide the halting problem for deterministic particle methods under the conditions in Eqs.~(\ref{eq:HaltingProblem1:conditionFiniteProperties})--(\ref{eq:HaltingProblem1:conditionNoParticleProduction}). 
	
\end{proof}

\section{Conclusion}
We used tools from automata theory to prove three theorems about the computational power of particle methods. Particle methods are a well-defined class of algorithms from scientific computing, used for computer simulations of both discrete and continuous models~\cite{Pahlke:2023} on sequential and parallel computers~\cite{Pahlke:2024}.
Despite their broad applicability, however, the theoretical limits of particle methods so far remained unclear. 
In the first two theorems, we have proven Turing powerfulness of particle methods under two non-trivial sets of restrictions, where one is not a sub-set of the other. This shows that while the most general particle method is trivially Turing powerful, particle methods remain Turing powerful even when severely resource-limited. Eventually, however, when restricting them beyond a certain limit, they become halting-decidable, as we have shown in the third theorem. This defines the border between Turing-powerful particle methods, and particle methods that are not Turing powerful any more.

The result that particle methods are Turing powerful, even if non-trivially restricted, shows the considerable theoretical potential of particle methods. Even though it does not tell us how useful a specific particle-method implementation of an algorithm is, we now know that one can express {\em any} Turing-computable function as a particle method with resource restrictions than enable practical implementation.

We have shown that particle methods are Turing powerful for mutually non-containing sets of non-trivial restrictions. This indicates that there is no unique set of maximal restrictions, but rather multiple ways of restricting particle methods such that they cannot be further restricted without losing Turing powerfulness.
Here, we have shown one sufficient set of restrictions leading to halting decidability. This provides us with an idea of when particle methods are not Turing powerful any more. 
However, we did not prove that the restrictions we considered are tight.
Hence, there might exist more restrictive sets for Turing powerfulness and less restrictive ones for halting decidability.

Finding more restrictive constraints under which particle methods are still Turing powerful, and less restrictive ones where particle methods become halting-decidable, are topics future work could address. For halting decidability, future work could also investigate more practically relevant constraints for actual software implementations. For example, many particle methods use counters with fixed limits to determine when to stop. This could be a starting point for more practically relevant constraints.

Overall, the present work on Turing powerfulness and halting decidability of particle methods opens up a discussion about the theoretical capabilities and limits of the particle-methods definition, and of numerical algorithms from scientific computing in general.

\section*{Acknowledgments}
We thank Udo Hebisch, TU Bergakademie Freiberg, for discussions at the beginning of this project. This work was funded in parts by the German Research Foundation (Deutsche Forschungsgemeinschaft, DFG) within the Research Training Group ``Role-based software infrastructures for continuous-context-sensitive systems'' (GRK 1907).

\newpage
\appendix
\section{Proof Details of the Turing Powerfulness of Particle Methods Under the First Set of Constraints}\label{app:A}
We provide the details for the proof of the Turing powerfulness of particle methods under the first set of constraints from Section \ref{sec:first}. The details include the proof that the particle method given in the main text indeed fulfills the constrains and is able to emulate an arbitrary Turing machine. 

\subsubsection*{\em Particle Method for Turing Machine Fulfills Constrains}
We prove that the constraints are fulfilled by the particle method.
The fulfillment of the constraints  (\ref{eq:TMsimple:condition:u}) and (\ref{eq:TMsimple:condition:i}) are directly visible in the definition of the particle method algorithm (\ref{eq:TMsimple:u}) and (\ref{eq:TMsimple:i}).

\begin{proof}
To prove that the evolve function is order-independent regarding $g$ (\ref{eq:TMsimple:condition:e:g}), we have to distinguish between four cases.

First, $\mathfrak k'=\mathfrak k''=\mathfrak m$
\begin{align}
	&{}_{1}e\left({}_1 e\left(g,p'\right),p''\right)\\
	&={}_{1}e\left(\left(\mathfrak q,\max( \Delta\mathfrak q,\overline{\mathfrak q}') ,\max(\mathfrak  d,\overline{\mathfrak d}') ,\mathfrak m,\mathfrak M\right),p''\right)\\
	&=\left(\mathfrak q,\max\left(\max(\Delta\mathfrak q,\overline{\mathfrak q}'),\overline{\mathfrak q}''\right) ,\max\left(\max(\mathfrak  d,\overline{\mathfrak d}'),\overline{\mathfrak d}''\right) ,\mathfrak m,\mathfrak M\right)\\
	&=\left(\mathfrak q,\max\left(\Delta\mathfrak q,\overline{\mathfrak q}',\overline{\mathfrak q}''\right) ,\max\left( \mathfrak d,\overline{\mathfrak d}',\overline{\mathfrak d}''\right) ,\mathfrak m,\mathfrak M\right)\\
	&=\left(\mathfrak q,\max\left(\max(\Delta\mathfrak q,\overline{\mathfrak q}''),\overline{\mathfrak q}'\right) ,\max\left(\max( \mathfrak d,\overline{\mathfrak d}''),\overline{\mathfrak d}'\right) ,\mathfrak m,\mathfrak M\right)\\
	&={}_{1}e\left(\left(\mathfrak q,\max( \Delta\mathfrak q,\overline{\mathfrak q}'') ,\max(\mathfrak d,\overline{\mathfrak d}'') ,\mathfrak m,\mathfrak M\right),p'\right)\\
	&=\underline{{}_{1}e\left({}_1 e\left(g,p''\right),p'\right)}.
\end{align}

Second, $\mathfrak k'=\mathfrak m$ and $\mathfrak k''\neq \mathfrak m$, then we set $\tilde{g}:={}_1 e\left(g,p'\right)$ and know $g={}_1 e\left(g,p''\right)$. From this follows
\begin{align}
	&{}_{1}e\left({}_1 e\left(g,p'\right),p''\right)\\
	&={}_{1}e\left(\tilde{g},p''\right)\\
	&=\tilde{g}\\
	&={}_1 e\left(g,p'\right)\\
	&=\underline{{}_{1}e\left({}_1 e\left(g,p''\right),p'\right)}.
\end{align}

Third, $\mathfrak k'\neq \mathfrak m$ and $\mathfrak k''= \mathfrak m$, then we know $g={}_1 e\left(g,p'\right)$ and set $\tilde{g}:={}_1 e\left(g,p''\right)$. From this follows
\begin{align}
	&{}_{1}e\left({}_1 e\left(g,p'\right),p''\right)\\
	&={}_{1}e\left(g,p''\right)\\
	&=\tilde{g}\\
	&={}_1 e\left(\tilde{g},p'\right)\\
	&=\underline{{}_{1}e\left({}_1 e\left(g,p''\right),p'\right)}.
\end{align}

Fourth, $\mathfrak k'\neq \mathfrak m$ and $\mathfrak k''\neq \mathfrak m$, then we know $g={}_1 e\left(g,p'\right)={}_1 e\left(g,p''\right)$. Hence,
\begin{align}
	&{}_{1}e\left({}_1 e\left(g,p'\right),p''\right)\\
	&={}_{1}e\left(g,p''\right)\\
	&=g\\
	&={}_1 e\left(g,p'\right)\\
	&=\underline{{}_{1}e\left({}_1 e\left(g,p''\right),p'\right)}.
\end{align}

Next, we prove the condition that the evolve function is independent of previous evolutions regarding $p$ (\ref{eq:TMsimple:condition:e:p}). We have to distinguish between two cases.

First, $\mathfrak k''=\mathfrak m$, for the evolve function, only the global variable's properties $\mathfrak q$, $\mathfrak m$, and $\mathfrak M$ are relevant. The evolve function does not change these properties. Hence, 
\begin{align}
	&{}_{2}e\left({}_1 e\left(g,p''\right),p'\right)\\
	&={}_{2}e\left(\left(\mathfrak q,\max\left(\Delta\mathfrak q,\overline{\mathfrak q}''\right) ,\max\left(\mathfrak d,\overline{\mathfrak d}''\right) ,\mathfrak m,\mathfrak M\right),p'\right)\\	&={}_{2}e\left(\left(\mathfrak q,\Delta\mathfrak q,  \mathfrak d ,\mathfrak m,\mathfrak M\right),p'\right)\\
	&=\underline{{}_{2}e\left(g,p'\right)}.
\end{align}

Second, $\mathfrak k''\neq \mathfrak m$, then we know $g={}_1 e\left(g,p''\right)$. Hence,
\begin{align}
	&{}_{2}e\left({}_1 e\left(g,p''\right),p'\right)\\
	&=\underline{{}_{2}e\left(g,p'\right)}.
\end{align}

The last constraint we prove is that the size of the global variable $\varsigma_g$ and each particle $\varsigma_p$, the time complexity of the stop function $\tau_{f}$, of the evolve function $\tau_e$ and of the evolve function of the global variable $\tau_{\overset{\circ }{e}}$ are in $O(\log(|\mathbf{p}^t|))$.
The only potentially growing properties are the cell index $\mathfrak k$ of the particles, the head position $\mathfrak m$, and the number of particles $\mathfrak M$.
All three numbers only increase if a new particle is created, hence when the number of particles increases. This happens for $\mathfrak m+\overline{\mathfrak d}  >\mathfrak M$ in the evolve function and in the evolve function $e$ of the global variable $\overset{\circ }{e}$. In theory, more than one particle could be created per step, but the properties would still be only advanced by one. Therefore, $\varsigma_g, \varsigma_p \in {O}\left(\log(|\mathbf{p}^t|)\right)$. For a particle methods instance for a Turing machine, as defined in section \ref{sec:TuringMachineByKozen}, the creation of multiple particles per step would not happen.
The operations in $f,e,\overset{\circ }{e}$ are only comparisons, additions, and replacements. If we assume these operations are done digit-vise and we know that the numbers are  in ${O}\left(\log(|\mathbf{p}^t|)\right)$, we can follow 
the time complexity of the functions $f,e,\overset{\circ }{e}$ are also in ${O}\left(\log(|\mathbf{p}^t|)\right)$. Hence,
\begin{align}
	\varsigma_{g^t} ,
	\varsigma_{p_j^t} ,
	\tau_{f} ,
	\tau_{e}, 
	\tau_{\overset{\circ }{e}}
	\in {O}\left(\log(|\mathbf{p}^t|)\right) .
\end{align}

Therefore the particle method fulfills all constraints from (\ref{eq:TMsimple:condition:u}) to (\ref{eq:TMsimple:condition:complexity}).

\end{proof}

\subsubsection*{\em Particle Method Emulates Arbitrary Turing Machine}
To prove that the particle method emulates an arbitrary Turing machine, we need to translate the start configuration of a Turing machine (\ref{eq:TuringMachineByKozen:startConfiguration}) into a particle methods instance. Then we prove for all states of the particle method that the back translation is the corresponding configuration of the Turing machine. The last step is to show that the particle method stops when the Turing machine reaches an accept or reject state.\\

\begin{proof}

We define for this particle method the translation function as
\begin{equation}\label{eq:TMsimple:psi}
	\psi \left(\left( q,\mathbf{x}\blank^{\omega } ,n\right)\right) :=[( q,\textbf{start},-1,n,|\mathbf{x} |) ,(( 1,x_{1}) ,( 2,x_{2}) ,...,(| \mathbf{x}| ,x_{|\mathbf{x} |}))],
\end{equation}

and the back translation as
\begin{equation}
	\psi ^{-1}\left(\left[\left(\mathfrak q,\Delta\mathfrak q ,\mathfrak d,\mathfrak m,\mathfrak M\right) ,(( \mathfrak k_{1} ,\mathfrak z_{1}) ,( \mathfrak k_{2} ,\mathfrak z_{2}) ,...,( \mathfrak k_{|\mathbf{p} |} ,\mathfrak z_{|\mathbf{p} |}))\right]\right)
	:=\left(\mathfrak q,\mathfrak z_{1} \mathfrak z_{2} \cdots \mathfrak z_{|\mathbf{p} |}{}\blank^{\omega } ,\mathfrak m\right).
\end{equation}

We see these translation functions are only copying values. The only part which might be a calculation is the length of the finite input string $|\mathbf x|$. Therefore, the translations do not carry any calculation of the Turing machine.

The instance of the particle method is defined by translating the start configuration $$\alpha^1=(\textbf{start},\mathbf{x}^1\blank^{\omega},1).$$ Hence,
\begin{equation}
	[g^1,\mathbf{p}^1]:=\psi(\alpha^1).
\end{equation}
We need additional criteria to be fulfilled by each particle methods step so that the particle method works. In total, the following criteria are needed:
\begin{equation}\label{eq:TMsimple:inductionStepConditions}
	\begin{array}{ l }
		\psi ^{-1}\left(\left[ g^{t} ,\mathbf{p}^{t}\right]\right) =\alpha ^{t}\\
		\land\ \mathfrak{M}^{t} =|\mathbf{p}^{t} |\\
		\land\ \mathfrak{m}^{t} \leq \mathfrak{M}^{t}\\
		\land\ \mathfrak{d}^{t} =-1\\
		\land\ \Delta \mathfrak{q}^{t} =\textbf {start}\\
		\land\ \forall p_{j}^{t} \in \mathbf{p}^{t} :\mathfrak{k}_{j}^{t} =j.
	\end{array}
\end{equation}

We prove through induction that the back translation of all particle method states is the Turing machine's corresponding configuration and that the criteria (\ref{eq:TMsimple:inductionStepConditions}) are fulfilled.

The base case regards the instance and the start configuration.
\begin{align}
	\label{eq:TMsimple:baseCase:start}
	&\psi ^{-1}\left( [g^1,\mathbf{p}^1]\right)\\
	&=\psi ^{-1}\left( \psi \left(\left( \textbf{start},\mathbf{x}^1\blank^{\omega } ,1\right)\right)\right)\\
	\label{eq:TMsimple:instance}
	&=\psi ^{-1}([( \textbf{start},\textbf{start},-1,1,|\mathbf{x}^1 |) ,(( 1,\vdash) ,( 2,x^1_{2}) ,...,(| \mathbf{x}^1| ,x^1_{|\mathbf{x}^1 |}))])\\
	&=\left( \textbf{start},\vdash x^1_{2} \cdots x^1_{|\mathbf{x}^1 |}{}\blank^{\omega } ,1\right)\\
	\label{eq:TMsimple:baseCase:end}
	&=\left( \textbf{start},\mathbf{x}^1\blank^{\omega } ,1\right)\\
	&=\underline{\alpha^1}.
\end{align}
The rest of the criteria (\ref{eq:TMsimple:inductionStepConditions}) follow directly from (\ref{eq:TMsimple:instance}).

For the induction step, we need to prove that under the condition that the criteria (\ref{eq:TMsimple:inductionStepConditions}) are fulfilled by $\left[ g^{t},\mathbf{p}^{t}\right]$ follows that the criteria (\ref{eq:TMsimple:inductionStepConditions}) are fulfilled by $\left[ g^{t+1},\mathbf{p}^{t+1}\right]$.

We start by calculation $\left[ g^{t+1} ,\mathbf{p}^{t+1}\right]=s\left(\left[ g^{t} ,\mathbf{p}^{t}\right]\right)$. Since the neighborhood is empty, there is no interaction. Hence, the state transition step is the evolve function $e$ and the evolve function of the global variable $\overset{\circ }{e}$. Therefore, we can rewrite $s$ to
\begin{equation}
	s\left(\left[ g^{t} ,\mathbf{p}^{t}\right]\right) =\left[\overset{\circ }{e}\left( g*_{_{1} e}\left( p_{1}^{t} ,...,p_{|\mathbf{p}^{t} |}^{t}\right)\right) ,
	{}_{2} e\left( p_{1}^{t}\right) \circ ...\circ {}_{2} e\left( p_{|\mathbf{p}^{t} |}^{t}\right)\right].
\end{equation}
We calculate the part of the global variable separately from that of the particle tuple. 
The evolve function is different from the identity just for $n^t=\mathfrak k_j$. This is only the case for the particle $p_{n^t}$. Hence, 
\begin{align}
	&g*_{_{1} e}\left( p_{1}^{t} ,...,p_{|\mathbf{p}^{t} |}^{t}\right)\\
	& =g*_{_{1} e}\left( p_{1}^{t} ,...,p_{n^{t} -1}^{t} ,p_{n^{t}}^{t} ,p_{n^{t} +1}^{t} ,...,p_{|\mathbf{p}^{t} |}^{t}\right)\\
	& =g*_{_{1} e}\left( p_{n^{t}}^{t} ,p_{n^{t} +1}^{t} ,...,p_{|\mathbf{p}^{t} |}^{t}\right)\\
	& =\left( q^{t} ,\max (\textbf{start},\overline{\mathfrak{q}}^{t} ),\max(-1,\overline{\mathfrak{d}}^{t} ),n^{t} ,|\mathbf{p}^{t} |\right) *_{_{1} e}\left( ,p_{n^{t} +1}^{t} ,...,p_{|\mathbf{p}^{t} |}^{t}\right)\\
	& =\left( q^{t} ,\overline{\mathfrak{q}}^{t} ,\overline{\mathfrak{d}}^{t} ,n^{t} ,|\mathbf{p}^{t} |\right) *_{_{1} e}\left( ,p_{n^{t} +1}^{t} ,...,p_{|\mathbf{p}^{t} |}^{t}\right)\\
	& =\left( q^{t} ,\overline{\mathfrak{q}}^{t} ,\overline{\mathfrak{d}}^{t} ,n^{t} ,|\mathbf{p}^{t} |\right)\\
	& =\left( q^{t} ,q^{t+1} ,\overline{\mathfrak{d}}^{t} ,n^{t} ,|\mathbf{p}^{t} |\right)
\end{align}
The evolve function of the global variable $\overset{\circ }{e}$ has two cases. First, $\mathfrak m+\mathfrak d > \mathfrak M$ in this case this means $n^t+{\mathfrak{d}}^{t}>|\mathbf{p}^{t}|$. Hence,
\begin{align}
	\overset{\circ }{e}\left( g*_{_{1} e}\left( p_{1}^{t} ,...,p_{|\mathbf{p}^{t} |}^{t}\right)\right) & =\overset{\circ }{e}\left(\left( q^{t} ,q^{t+1} ,{\mathfrak{d}}^{t} ,n^{t} ,|\mathbf{p}^{t} |\right)\right)\\
	& =\left( q^{t+1} ,\textbf{start},-1,n^{t} +{\mathfrak{d}}^{t} ,|\mathbf{p}^{t} |+1\right)\\
	& =\underline{\left( q^{t+1} ,\textbf{start},-1,n^{t+1} ,|\mathbf{p}^{t} |+1\right)}.
\end{align}
Second, else. Hence,
\begin{align}
	&\overset{\circ }{e}\left( g*_{_{1} e}\left( p_{1}^{t} ,...,p_{|\mathbf{p}^{t} |}^{t}\right)\right)\\
	& =\overset{\circ }{e}\left(\left( q^{t} ,q^{t+1} ,{\mathfrak{d}}^{t} ,n^{t} ,|\mathbf{p}^{t} |\right)\right)\\
	& =\left( q^{t+1} ,\textbf{start},-1,\max\left( 1,n^{t} +{\mathfrak{d}}^{t}\right) ,|\mathbf{p}^{t} |\right)
\end{align}
We know $n^t\geq 1$, $p_1^t=(1,\vdash)$, and
$\delta(q,\vdash)=(\overline q,\vdash,1)$. 
Therefore, if $n^t=1$, then $\mathfrak{d}^{t}=1$. Hence,
\begin{align}
	&=\left( q^{t+1} ,\textbf{start},-1,n^{t} +{\mathfrak{d}}^{t} ,|\mathbf{p}^{t} |\right)\\
	&=\underline{\left( q^{t+1} ,\textbf{start},-1,n^{t+1},|\mathbf{p}^{t} |\right)}.
\end{align}
The global variable is finished. Next is the particle tuple.
We know the evolve function differs from the identity for $n^t=\mathfrak k_j$. This is the case only for the particle $p_{n^t}$.
We need to distinguish two cases:

First, when $\mathfrak m+\overline{\mathfrak d} > \mathfrak M$ in this case $n^t+\overline{\mathfrak d}^t > |\mathbf{p}^{t} |$.  Additionally, we know that $n^t\leq |\mathbf{p}^{t} |$ and $\overline{\mathfrak d}^t\in\{-1,1\}$. This implies that $n^t=|\mathbf{p}^{t} |$ and $\overline{\mathfrak d}^t=1$. Hence,
\begin{align}
	& {}_{2} e\left( p_{1}^{t}\right) \circ ...\circ {}_{2} e\left( p_{|\mathbf{p}^{t} |}^{t}\right)\\
	& =\left( p_{1}^{t}\right) \circ ...\circ \left( p_{|\mathbf{p}^{t} |-1}^{t}\right) \circ \left(\left( |\mathbf{p}^{t} |,\overline{\mathfrak{z}}^{t}\right) ,\left( |\mathbf{p}^{t} |+1,_{-}\right)\right)\\
	& =\underline{\left( p_{1}^{t} ,...,p_{|\mathbf{p}^{t} |-1}^{t} ,\left( |\mathbf{p}^{t} |,\overline{\mathfrak{z}}^{t}\right) ,\left( |\mathbf{p}^{t} |+1,_{-}\right)\right)}
\end{align}
Second, in the case $n^t+\overline{\mathfrak d}^t \leq |\mathbf{p}^{t} |$, we get
\begin{align}
	& {}_{2} e\left( p_{1}^{t}\right) \circ ...\circ {}_{2} e\left( p_{|\mathbf{p}^{t} |}^{t}\right)\\
	& =\left( p_{1}^{t}\right) \circ ...\circ \left( p_{n^{t} -1}^{t}\right) \circ \left(\left( n^{t} ,\overline{\mathfrak{z}}^{t}\right)\right) \circ \left( p_{n^{t} +1}^{t}\right) \circ ...\circ \left( p_{|\mathbf{p}^{t} |}^{t}\right)\\
	& =\underline{\left( p_{1}^{t} ,...,p_{n^{t} -1}^{t} ,\left( n^{t} ,\overline{\mathfrak{z}}^{t}\right) ,p_{n^{t} +1}^{t} ,...,p_{|\mathbf{p}^{t} |}^{t}\right)}.
\end{align}
Using these results, we prove that the criteria  (\ref{eq:TMsimple:inductionStepConditions}) hold. 
First, for $n^t+\overline{\mathfrak d}^t \geq |\mathbf{p}^{t} |$.
\begin{align}
	& \psi ^{-1}\left(\left[\left( q^{t+1} ,\textbf{start},-1,n^{t+1} ,|\mathbf{p}^{t} |+1\right) ,\left( p_{1}^{t} ,...,p_{|\mathbf{p}^{t} |-1}^{t} ,\left( |\mathbf{p}^{t} |,\overline{\mathfrak{z}}^{t}\right) ,\left( |\mathbf{p}^{t} |+1,_{-}\right)\right)\right]\right)\\
	& =\left( q^{t+1} ,x_{1}^{t} x_{2}^{t} \cdots x_{|\mathbf{p}^{t} |-1}^{t}\overline{\mathfrak{z}}^{t}{}_{-}{}{_{-}}^{\omega } ,n^{t+1}\right)\\
	& =\left( q^{t+1} ,x_{1}^{t} x_{2}^{t} \cdots x_{|\mathbf{p}^{t} |-1}^{t}\overline{\mathfrak{z}}^{t}{}{_{-}}^{\omega +1} ,n^{t+1}\right)\\
	& =\left( q^{t+1} ,x_{1}^{t} x_{2}^{t} \cdots x_{|\mathbf{p}^{t} |-1}^{t}\overline{\mathfrak{z}}^{t}{}{_{-}}^{\omega } ,n^{t+1}\right)\\
	& =\underline{\alpha ^{t+1}}
\end{align}
There was a particle added to $\mathbf{p}^t$. The length increased by one, and so did 
\begin{equation}
	\underline{\mathfrak{M}^{t+1}}=|\mathbf{p}^t|+1\underline{=|\mathbf{p}^{t+1}|}.
\end{equation}
From 
$\overline{\mathfrak d}^t=1$,
$\mathfrak{m}^{t} \leq \mathfrak{M}^{t}$ we  can follow
\begin{align}
	\mathfrak{m}^{t} & \leq \mathfrak{M}^{t}\\
	\rightarrow \ n^{t} & \leq |\mathbf{p}^{t} |\\
	\rightarrow \ n^{t} +1 & \leq |\mathbf{p}^{t} |+1\\
	\rightarrow \ n^{t+1} & \leq |\mathbf{p}^{t+1} |\\
	\rightarrow\ \underline{\mathfrak{m}^{t+1}} & \underline{\leq \mathfrak{M}^{t+1}}.
\end{align}
We directly see that 
\begin{equation}
	\mathfrak{d}^{t+1}=-1, \quad \Delta \mathfrak{q}^{t+1}=\textbf{start}.
\end{equation}
The particles are not change except for the particle $p_{|\mathbf{p}^{t} |}$, where the symbol $\mathfrak{z}_{|\mathbf{p}^{t}|}$ changed  and the index $\mathfrak{k}_{|\mathbf{p}^{t}|}$stayed the same.
But a new particle was added at the end, i.e., at the index $|\mathbf{p}^{t}|+1$, and it got the index $\mathfrak{k}_{|\mathbf{p}^{t}|+1}=|\mathbf{p}^{t}|+1$. Hence, we can follow
\begin{equation}
	\forall p_{j}^{t+1} \in \mathbf{p}^{t+1} :\mathfrak k_{j}^{t+1} =j.
\end{equation}
Second for $n^t+\overline{\mathfrak d}^t \leq |\mathbf{p}^{t} |$:
\begin{align}
	&\psi ^{-1}\left(\left[\left( q^{t+1} ,\textbf{start},-1,n^{t+1} ,|\mathbf{p}^{t} |\right) ,\left( p_{1}^{t} ,...,p_{n^{t} -1}^{t} ,\left( n^{t} ,\overline{\mathfrak{z}}^{t}\right) ,p_{n^{t} +1}^{t} ,...,p_{|\mathbf{p}^{t} |}^{t}\right)\right]\right)\\
	&=\left( q^{t+1} ,x_{1}^{t} x_{2}^{t} \cdots x_{n^{t} -1}^{t}\overline{\mathfrak{z}}^{t} x_{n^{t} +1}^{t} \cdots x_{|\mathbf{p}^{t} |}^{t} \ {_{-}}^{\omega } ,n^{t+1}\right)\\
	&=\left( q^{t+1} ,\mathbf{x}^{t+1} \ {_{-}}^{\omega } ,n^{t+1}\right)\\
	&=\underline{\alpha ^{t+1}}.
\end{align}
Since there was no particle added to $\mathbf{p}^t$, the length stays the same
\begin{equation}
	\underline{\mathfrak{M}^{t+1}}=\mathfrak{M}^{t}=|\mathbf{p}^t|\underline{=|\mathbf{p}^{t+1}|}.
\end{equation}
We can also follow
\begin{equation}
	\underline{\mathfrak{m}^{t+1}}=n^{t+1}=n^t+\overline{\mathfrak d}^t\ \ \underline{\leq}\ \ |\mathbf{p}^{t} |=|\mathbf{p}^{t+1} |=\underline{\mathfrak{M}^{t+1}},
\end{equation}
and directly see that 
\begin{equation}
	\mathfrak{d}^{t+1}=-1, \quad \Delta \mathfrak{q}^{t+1}=\textbf{start}.
\end{equation}
The particles are not changed except for the particle $p_{n^t}$, and the index $\mathfrak{k}_{n^t}$ was not changed. Hence,
\begin{equation}
	\forall p_{j}^{t} \in \mathbf{p}^{t} :\mathfrak k_{j}^{t} =j \ \rightarrow \  \forall p_{j}^{t+1} \in \mathbf{p}^{t+1} :\mathfrak k_{j}^{t+1} =j.
\end{equation}
This completes the induction step.

The last part to prove is that the particle method stops if and only if the Turing machine holds.
\begin{equation}
	f(g^t)=\top \ \leftrightarrow \ q^t\in \{\textbf{accept}, \textbf{reject}\}
\end{equation} 
This is proven by
\begin{align}
	& f\left( g^{t}\right) =\top \\
	\leftrightarrow  & \mathfrak{q}^{t} \in \{\textbf{accept},\ \textbf{reject}\}\\
	\xleftrightarrow{\psi ^{-1}\left(\left[ g^{t} ,\mathbf{p}^{t}\right]\right) =\alpha ^{t}} & q^{t} \in \{\textbf{accept},\ \textbf{reject}\}.
\end{align}
\end{proof}

\section{Proof Details of the Turing Powerfulness of Particle Methods Under the Second Set of Constraints}\label{app:B}
We provide the details for the proof of the Turing powerfulness of particle methods under the second set of constraints from Section \ref{sec:second}. The details include the proof that the particle method given in the main text indeed fulfills the constrains and is able to emulate an arbitrary Turing machine.

\subsubsection*{\em Particle Method of Turing Machine Fulfills Constrains}
We prove that the particle method fulfills all constraints from (\ref{eq:TMcomplex:condition:PG}) to (\ref{eq:TMcomplex:condition:functions:spacecomplexity}).
\begin{proof}
The condition that $P$ and $G$ are finite (\ref{eq:TMcomplex:condition:PG})
is fulfilled since the Cartesian product of two finite sets is a finite set. All sets are finite, inducing $\Gamma$ and $Q$. Hence,  $|P|<\infty$ 	and $ |G|<\infty$.

The interact function $i$ is defined so that the second particle stays the same for each case. Therefore it is a pull interaction (\ref{eq:TMcomplex:condition:i:pull}).

The interact function $i$ can only change $\Delta \mathfrak{h}, \Delta \mathfrak{o}, \mathfrak{a}$ and only depends on $\mathfrak{z},\mathfrak{h}, \mathfrak{o}$ from the second particle, i.e., it depends only on properties that are not changed during the interactions. Hence,
\begin{align}
	_{1} i_{g} (p_{j} ,{}_{1} i_{g} (p_{k} ,p_{k'} ))
	&={}_{1} i_{g} (p_{j} ,(\mathfrak{z}_k,\mathfrak{h}_k ,\Delta \mathfrak{h}'_k  ,\mathfrak{o}_k ,\Delta \mathfrak{o}'_k ,\mathfrak{a}'_k) )\\
	&={}_{1} i_{g} (p_{j} ,(\mathfrak{z}_k,\mathfrak{h}_k ,\Delta \mathfrak{h}_k ,\mathfrak{o}_k ,\Delta \mathfrak{o}_k ,\mathfrak{a}_k) )\\
	&=\underline{{}_{1} i_{g} (p_{j} ,p_{k} )}.
\end{align}
Therefore, $i$ is independent of previous interactions (\ref{eq:TMcomplex:condition:i:Previous}).

We know that the interact function $i$ does not change any property of $p_j$ and $p_k$ that is used to change properties in the interact function $i$. Therefore, it is sufficient to focus on how the properties are changed to prove the order independence of $i$ (\ref{eq:TMcomplex:condition:i:Order}). For readability, we leaf out the conditions they can be handled similarly as in (\ref{eq:TMcomplex:condition:e:order:proof:start}) to (\ref{eq:TMcomplex:condition:e:order:proof:end}).

\begin{align}
	& _{1} i(_{1} i(g,p_{j} ,p_{k} ),p_{k'} )\\
	& ={}_{1} i\left(\left(\mathfrak{z}_{j} ,\mathfrak{h}_{j} ,\begin{cases}
		\Delta \mathfrak{h}_{j}\\
		1
	\end{cases} ,\Delta \mathfrak{h}_{j} ,\mathfrak{o}_{j} ,\begin{cases}
		\Delta \mathfrak{o}_{j}\\
		\max (\Delta \mathfrak{o}_{j} ,\overline{\mathfrak{d}}_{k} )
	\end{cases} ,\begin{cases}
		\mathfrak{o}_{j}\\
		1
	\end{cases}\right) ,p_{k'}\right)\\
	& =\left(\mathfrak{z}_{j} ,\mathfrak{h}_{j} ,\begin{cases}
		\begin{cases}
			\Delta \mathfrak{h}_{j}\\
			1
		\end{cases}\\
		1
	\end{cases} ,\Delta \mathfrak{h}_{j} ,\mathfrak{o}_{j} ,\begin{cases}
		\begin{cases}
			\Delta \mathfrak{o}_{j}\\
			\max (\Delta \mathfrak{o}_{j} ,\overline{\mathfrak{d}}_{k} )
		\end{cases}\\
		\max\left(\begin{cases}
			\Delta \mathfrak{o}_{j}\\
			\max (\Delta \mathfrak{o}_{j} ,\overline{\mathfrak{d}}_{k} )
		\end{cases} ,\overline{\mathfrak{d}}_{k'}\right)
	\end{cases} ,\begin{cases}
		\begin{cases}
			\mathfrak{o}_{j}\\
			1
		\end{cases}\\
		1
	\end{cases}\hspace{-7mm}\right)\\
	& =\left(\mathfrak{z}_{j} ,\mathfrak{h}_{j} ,\begin{cases}
		\begin{cases}
			\Delta \mathfrak{h}_{j}\\
			1
		\end{cases}\\
		1
	\end{cases} ,\Delta \mathfrak{h}_{j} ,\mathfrak{o}_{j} ,\begin{cases}
		\Delta \mathfrak{o}_{j}\\
		\max (\Delta \mathfrak{o}_{j} ,\overline{\mathfrak{d}}_{k} )\\
		\max (\Delta \mathfrak{o}_{j} ,\overline{\mathfrak{d}}_{k'} )\\
		\max (\Delta \mathfrak{o}_{j} ,\overline{\mathfrak{d}}_{k} ,\overline{\mathfrak{d}}_{k'} )
	\end{cases} ,\begin{cases}
		\begin{cases}
			\mathfrak{o}_{j}\\
			1
		\end{cases}\\
		1
	\end{cases}\right)\\
	& =\left(\mathfrak{z}_{j} ,\mathfrak{h}_{j} ,\begin{cases}
		\begin{cases}
			\Delta \mathfrak{h}_{j}\\
			1
		\end{cases}\\
		1
	\end{cases} ,\Delta \mathfrak{h}_{j} ,\mathfrak{o}_{j} ,\begin{cases}
		\begin{cases}
			\Delta \mathfrak{o}_{j}\\
			\max (\Delta \mathfrak{o}_{j} ,\overline{\mathfrak{d}}_{k'} )
		\end{cases}\\
		\max\left(\begin{cases}
			\Delta \mathfrak{o}_{j}\\
			\max (\Delta \mathfrak{o}_{j} ,\overline{\mathfrak{d}}_{k'} )
		\end{cases} ,\overline{\mathfrak{d}}_{k}\right)
	\end{cases} ,\begin{cases}
		\begin{cases}
			\mathfrak{o}_{j}\\
			1
		\end{cases}\\
		1
	\end{cases}\hspace{-7mm}\right)\\
	& ={}_{1} i\left(\left(\mathfrak{z}_{j} ,\mathfrak{h}_{j} ,\begin{cases}
		\Delta \mathfrak{h}_{j}\\
		1
	\end{cases} ,\Delta \mathfrak{h}_{j} ,\mathfrak{o}_{j} ,\begin{cases}
		\Delta \mathfrak{o}_{j}\\
		\max (\Delta \mathfrak{o}_{j} ,\overline{\mathfrak{d}}_{k'} )
	\end{cases} ,\begin{cases}
		\mathfrak{o}_{j}\\
		1
	\end{cases}\right) ,p_{k}\right)\\
	& =\underline{_{1} i(_{1} i(g,p_{j} ,p_{k'} ),p_{k} )}
\end{align}

We prove that the evolve function $e$ is order-independent regarding $g$ (\ref{eq:TMcomplex:condition:e:order}) by
\begin{align}
	\label{eq:TMcomplex:condition:e:order:proof:start}
	& _{1} e(_{1} e(g,p'),p'')\\
	& ={}_{1} e\left(\begin{cases}
		(\mathfrak{q} ,\max (\Delta \mathfrak{q} ,\overline{\mathfrak{q}} ')) & \text{if }\mathfrak{h} '=1\\
		g & \text{else}
	\end{cases} ,p''\right)\\
	& =\begin{cases}
		\begin{cases}
			(\mathfrak{q} ,\max (\max (\Delta \mathfrak{q} ,\overline{\mathfrak{q}} '),\overline{\mathfrak{q}} '') & \text{if }\mathfrak{h} '=1\\
			(\mathfrak{q} ,\max (\Delta \mathfrak{q} ,\overline{\mathfrak{q}} '')) & \text{else}
		\end{cases} & \text{if }\mathfrak{h} ''=1\\
		\begin{cases}
			(\mathfrak{q} ,\max (\Delta \mathfrak{q} ,\overline{\mathfrak{q}} ')) & \text{if }\mathfrak{h} '=1\\
			g & \text{else}
		\end{cases} & \text{else}
	\end{cases}\\
	& =\begin{cases}
		(\mathfrak{q} ,\max (\Delta \mathfrak{q} ,\overline{\mathfrak{q}} ',\overline{\mathfrak{q}} '') & \text{if }\mathfrak{h} '=1\ \land \ \mathfrak{h} ''=1\\
		(\mathfrak{q} ,\max (\Delta \mathfrak{q} ,\overline{\mathfrak{q}} '')) & \text{if }\mathfrak{h} '\neq 1\land \ \mathfrak{h} ''=1\\
		(\mathfrak{q} ,\max (\Delta \mathfrak{q} ,\overline{\mathfrak{q}} ')) & \text{if }\mathfrak{h} '=1\ \land \ \mathfrak{h} ''\neq 1\\
		g & \text{else}
	\end{cases}
\end{align}
\begin{align}
	& =\begin{cases}
		\begin{cases}
			(\mathfrak{q} ,\max (\max (\Delta \mathfrak{q} ,\overline{\mathfrak{q}} '),\overline{\mathfrak{q}} '') & \text{if }\mathfrak{h} ''=1\\
			(\mathfrak{q} ,\max (\Delta \mathfrak{q} ,\overline{\mathfrak{q}} ')) & \text{else}
		\end{cases} & \text{if }\mathfrak{h} '=1\\
		\begin{cases}
			(\mathfrak{q} ,\max (\Delta \mathfrak{q} ,\overline{\mathfrak{q}} '')) & \text{if }\mathfrak{h} ''=1\\
			g & \text{else}
		\end{cases} & \text{else}
	\end{cases}\\
	& ={}_{1} e\left(\begin{cases}
		(\mathfrak{q} ,\max (\Delta \mathfrak{q} ,\overline{\mathfrak{q}} '')) & \text{if }\mathfrak{h} ''=1\\
		g & \text{else}
	\end{cases} ,p'\right)\\
	& \underline{={}_{1} e(_{1} e(g,p''),p')} .
	\label{eq:TMcomplex:condition:e:order:proof:end}
\end{align}

The evolve function changes only $\Delta \mathfrak{q}$ in the global variable, but changes by the evolve function only depend on the property $\mathfrak{q}$ of the global variable, i.e., in the series of evolutions only properties are changed that do not influence the change of following evolutions, hence the evolve function is independent of previous evolutions (\ref{eq:TMcomplex:condition:e:previous})
\begin{align}
	_{2} e(_{1} e( g,p'') ,p') & ={}_{2} e((\mathfrak{q} ,\Delta \mathfrak{q} '') ,p')\\
	& ={} _{2} e((\mathfrak{q} ,\Delta \mathfrak{q}) ,p')\\
	& =\underline{_{2} e( g,p')}.
\end{align}

The neighborhood function $u$ only uses $|\mathbf{p}|$ and $j$ to determine the neighbor indices $k$. Hence,
\begin{equation} 
	u([g,\mathbf p], j)= (k\in \{1,...,|\mathbf{p}|\}:\Omega(j,k)=\top)
\end{equation}
with 
\begin{equation} 
	\Omega(j,k):=(k=j-1\lor k=j+1).
\end{equation}
Therefore, $u$ is independent of $g$ and the values of $p_j\in\mathbf{p}$
(\ref{eq:TMcomplex:condition:u:valueindependence}).

The interact function cannot add or destroy particles. Therefore it can not change the number of particles in a tuple $|\mathbf p *_{\iota^I_{(g,k')}} (k'')|=|\mathbf p|$. Hence,
\begin{equation} 
	u([g,\mathbf p *_{\iota^I_{(g,k')}} (k'')], j)=u([g,\mathbf p], j).
\end{equation}
Therefore, $u$ is independent of previous interactions (\ref{eq:TMcomplex:condition:u:previous}).

The functions $e,i,f, \overset{\circ}{e}$ 
consist of evaluations of basic comparisons, $\max$-functions and the transition function of the Turing machine
$\delta\subseteq Q\times\Gamma\times Q\times \Gamma\times \{-1,1\}$. 
There are no loops or equivalent. Since $G$, $P$, $Q$, and $\Gamma$ are finite so is $\delta$ and with it the time complexities of
$e,i,f, \overset{\circ }{e}$ (\ref{eq:TMcomplex:condition:functions:timecomplexity}) 
\begin{equation}
	\tau_e,\tau_i,\tau_f,\tau_{\overset{_\circ}{e}}\in O(1),
\end{equation}
and the space complexities of $e,i,u,f, \overset{\circ }{e}$ (\ref{eq:TMcomplex:condition:functions:spacecomplexity})
\begin{equation}
	\varsigma_e,\varsigma_i,\varsigma_u,\varsigma_f,\varsigma_{\overset{_\circ}{e}}\in O(1).
\end{equation}
The time complexity of the neighborhood function $\tau_u$ is in  ${O}(log(|\mathbf{p}^t|))$ if one considers $|\mathbf{p}^t|$ to be known.

\end{proof}

\subsubsection*{\em Particle Method Emulates Arbitrary Turing Machine}
To prove that the particle method emulates an arbitrary Turing machine, we need to translate the start configuration of a Turing machine (\ref{eq:TuringMachineByKozen:startConfiguration}) into a particle methods instance. Then we prove for all states of the particle method that the back translation is the corresponding configuration of the Turing machine. The last step is to show that the particle method stops when the Turing machine reaches an accept or reject state.

\begin{proof}
We define for this particle method the translation function as
\begin{align}\label{eq:TMcomplex:psi}
	&\psi \left(\left( q,\mathbf{x}\blank^{\omega } ,n\right)\right)\\
	& :=\left[ (\underbrace{q}_{\mathfrak{q}} ,\underbrace{start}_{\Delta \mathfrak{q}} ),\left(\begin{array}{ c c c c c c }
		(\underbrace{x_{1}}_{\mathfrak{z}}\mathfrak{,} & \underbrace{0}_{\mathfrak{h}} , & \underbrace{0}_{\Delta \mathfrak{h}} , & \underbrace{-1}_{\mathfrak{o}} , & \underbrace{-1}_{\Delta \mathfrak{o}} , & \underbrace{0}_{\mathfrak{a}})\\
		\vdots  &  &  &  &  & \\
		( x_{n-1} , & 0, & 0, & -1, & -1, & 0)\\
		( x_{n} , & 1, & 0, & -1, & -1, & 0)\\
		( x_{n+1} , & -1, & 0, & -1, & -1, & 0)\\
		\vdots  &  &  &  &  & \\
		( x_{|\mathbf{x} |} & 0, & 0, & -1, & -1, & 0)
	\end{array}\right)^{\mathbf{T}}\right],
\end{align}
and the back translation as
\begin{equation}
	\psi ^{-1}([(\mathfrak{q} ,\Delta \mathfrak{q}) ,( p_{1} ,...,p_{|\mathbf{p} |})]) :=\left(\mathfrak{q} ,\mathfrak z_{1} \mathfrak z_{2} \cdots \mathfrak{z}_{|\mathbf{p} |}{}\blank^{\omega },\ j:\mathfrak h_{j} =1\right)
\end{equation}
We see these translation functions are only copying values.  Therefore, the translations do not carry any calculation of the Turing machine.

The instance of the particle method is defined by translating the start configuration $$\alpha^1=(\textbf{start},\mathbf{x}^1\blank^{\omega},1).$$ Hence,
\begin{equation}
	[g^1,\mathbf{p}^1]:=\psi(\alpha^1).
\end{equation}
We need additional criteria to be fulfilled by each particle methods step so that the particle method works. In total, the following criteria are needed:
\begin{align}
	\begin{array}{ l }\label{eq:TMcomplex:inductionStepConditions}
		\psi ^{-1}\left(\left[ g^{t} ,\mathbf{p}^{t}\right]\right) =\alpha ^{t} =\left( q^{t} ,\mathbf{x}^{t} \_^{\omega } ,n^{t}\right)\\
		\land\ \mathfrak{h}_{n^{t}}^{t} =1\ \land \ \mathfrak{h}_{n^{t} -\mathfrak{o}_{n^{t}}^{t}}^{t} =-1_{\myh} \\
		\land\ \forall j\in \left\{1,...,|\mathbf{p}^{t} |\right\} \backslash \left\{n^{t} ,n^{t} -\mathfrak{o}_{n^{t}}^{t}\right\} :\ \mathfrak{h}_{j}^{t} =0\\
		\land\ \forall j\in \left\{1,...,|\mathbf{p}^{t} |\right\} :\Delta \mathfrak{h}_{j}^{t} =0\ \land \Delta \mathfrak{o}_{j}^{t} =-1\land \mathfrak{a}_{j}^{t} =0\\
		\land\ \Delta q^{t} =\textbf{start}.
	\end{array}
\end{align}

We prove through induction that the back translation of all particle method states is the Turing machine's corresponding configuration and that the criteria (\ref{eq:TMcomplex:inductionStepConditions}) are fulfilled.

The base case is for the particle method instance and the start configuration.
\begin{align}
	\label{eq:TMcomplex:baseCase:start}
	&\psi ^{-1}\left( [g^1,\mathbf{p}^1]\right)\\
	&=\psi ^{-1}\left( \psi \left(\left( \textbf{start},\mathbf{x}^1\blank^{\omega } ,1\right)\right)\right)\\
	\label{eq:TMcomplex:instance}
	& =\psi ^{-1}\left(\left[ (\underbrace{\textbf{start}}_{\mathfrak{q}} ,\underbrace{\textbf{start}}_{\Delta \mathfrak{q}} ),\left(\begin{array}{ c c c c c c }
		(\underbrace{\vdash}_{\mathfrak{z}}\mathfrak{,} & \underbrace{1}_{\mathfrak{h}} , & \underbrace{0}_{\Delta \mathfrak{h}} , & \underbrace{-1}_{\mathfrak{o}} , & \underbrace{-1}_{\Delta \mathfrak{o}} , & \underbrace{0}_{\mathfrak{a}})\\
		( x^1_{2} , & -1, & 0, & -1, & -1, & 0)\\
		( x^1_{3} , & 0, & 0, & -1, & -1, & 0)\\
		\vdots  &  &  &  &  & \\
		( x^1_{|\mathbf{x}^1 |} & 0, & 0, & -1, & -1, & 0)
	\end{array}\right)^{\mathbf{T}}\right]\right)\\
	& =\left( \textbf{start},\vdash x^1_{2} \cdots x^1_{|\mathbf{x}^1 |}{}\blank^{\omega } ,1\right)\\
	& =\left( \textbf{start},\mathbf{x}^1\blank^{\omega } ,1\right)\\
	&=\underline{\alpha^1}.
	\label{eq:TMcomplex:baseCase:end}
\end{align}
The rest of the criteria (\ref{eq:TMcomplex:inductionStepConditions}) follow directly from (\ref{eq:TMcomplex:instance}).

For the induction step, we need to prove that if the criteria (\ref{eq:TMcomplex:inductionStepConditions}) are valid for $[g^t,\mathbf{p}^t]$, it is also true for $[g^{t+1},\mathbf{p}^{t+1}]=s([g^t,\mathbf{p}^t])$.
We divide the induction step into six cases.

First, we prove it for the case that the head is on the first cell/ particle, $\mathfrak{h}^t_1=1$. We use the criteria (\ref{eq:TMcomplex:inductionStepConditions}) to set up $[g^t,\mathbf{p}^t]$ and calculate from it $[g^{t+1},\mathbf{p}^{t+1}]$.  Since $p^t_0$ does not exist, $\mathbf{h}^t_2=-1$ and $\mathfrak{o}^t_1=-1$. From this and the definition of the Turing machine especially (\ref{eq:defTM:delta:CellOneRule}) follows that 
\begin{equation}
	\forall t\in \{1,...,T\}:  \delta(\mathfrak{q}^t,\mathfrak{z}^t_1)=\delta(\mathfrak{q}^t,\vdash)=(\overline{\mathfrak{q}},\vdash,1 )=:(\overline{\mathfrak{q}}^t,\overline{\mathfrak{z}}^t_1,\overline{\mathfrak{d}}^t_1).
\end{equation}
We calculate the interaction of $p^t_1$ with its neighbors
\begin{align}
	\tilde{\mathbf{p}}^t
	:=&\ \intII_g(\mathbf{p}^t,1)\\
	=&\ \mathbf{p}^t *_{\intI_{(g^t,1)}} u([g^t,\mathbf{p}^t],1)\\
	=&\ \mathbf{p}^t *_{\intI_{(g^t,1)}} (2)\\
	=&\ \bigg(\Bigl(\underbrace{\vdash}_{\mathfrak{z}},\underbrace{1}_{\mathfrak{h}} ,\underbrace{0}_{\Delta \mathfrak{h}} ,\underbrace{-1}_{\mathfrak{o}} ,\underbrace{-1}_{\Delta \mathfrak{o}} ,\underbrace{1}_{\mathfrak{a}}\Bigr) ,p_{2}^{t} ,...,p_{|\mathbf{p}^{t} |}^{t}\bigg).
\end{align}
and the interactions of $p^t_2$ with its neighbors
\begin{align}
	\tilde{\tilde{\mathbf{p}}}^t
	:=&\ \intII_g(\tilde{\mathbf{p}}^t,2)\\
	=&\ \tilde{\mathbf{p}}^t *_{\intI_{(g^t,2)}} u([g^t,\mathbf{p}^t],2)\\
	=&\ \tilde{\mathbf{p}}^t *_{\intI_{(g^t,2)}} (1,3).
\end{align}
The particle $p_1$ is the only particle with $\mathfrak{h}^t_1=1$ and $p_2$ with $\mathfrak{h}^t_2=-1$. Hence, $\mathfrak{h}^t_3=0$ and the interact function is for $p_2$ with $p_3$ the identity ($i(g,p_2,p_3)=(p_2,p_3)$). Therefore,

\begin{align}
	\tilde{\tilde{\mathbf{p}}}^{t} & =\bigg(\Bigl(\mathfrak{\underbrace{\vdash }_{\mathfrak{z}} ,}\underbrace{1}_{\mathfrak{h}} ,\underbrace{0}_{\Delta \mathfrak{h}} ,\underbrace{-1}_{\mathfrak{o}} ,\underbrace{-1}_{\Delta \mathfrak{o}} ,\underbrace{1}_{\mathfrak{a}}\Bigr) ,\notag\\
	& \quad \Bigl(\underbrace{x_{2}^{t}}_{\mathfrak{z}} ,\underbrace{-1}_{\mathfrak{h}} ,\underbrace{1}_{\Delta \mathfrak{h}} ,\mathfrak{o}_{2} ,\underbrace{\max\left( -1,\overline{\mathfrak{d}}_{1}^{t}\right)}_{\Delta \mathfrak{o}} ,\underbrace{0}_{\mathfrak{a}}\Bigr) ,p_{3}^{t} ,...,p_{|\mathbf{p}^{t} |}^{t}\bigg)\\
	& =\bigg(\Bigl(\mathfrak{\underbrace{\vdash }_{\mathfrak{z}} ,}\underbrace{1}_{\mathfrak{h}} ,\underbrace{0}_{\Delta \mathfrak{h}} ,\underbrace{-1}_{\mathfrak{o}} ,\underbrace{-1}_{\Delta \mathfrak{o}} ,\underbrace{1}_{\mathfrak{a}}\Bigr) ,\notag\\
	& \ \quad \Bigl(\underbrace{x_{2}^{t}}_{\mathfrak{z}} ,\underbrace{-1}_{\mathfrak{h}} ,\underbrace{1}_{\Delta \mathfrak{h}} ,\mathfrak{o}_{2} ,\underbrace{1}_{\Delta \mathfrak{o}} ,\underbrace{0}_{\mathfrak{a}}\Bigr) ,p_{3}^{t} ,...,p_{|\mathbf{p}^{t} |}^{t}\bigg)
\end{align}
For all other particles, the interact function is the identity, only $\mathfrak{h}_1^t=1$ and $\forall j\in \{3,...,|\mathbf{p}^t|\}: 1\notin u([g^t,\mathbf{p}^t],j)$. Hence,
\begin{equation}
	\intIII([g^t,\mathbf{p}^t])=\tilde{\tilde{\mathbf{p}}}^t.
\end{equation}
Next is the evolution step. The evolve function $e$ is the identity for the particles except if $\mathfrak{h}=1$ this is only true for $p^t_1$ or if $\Delta\mathfrak{h}=1$ this is only true for $p^t_2$ or if $\mathfrak{h}=-1$ only true for $p^t_2$. $e$ is also the identity for the global variable except if $\mathfrak{h}=1$. Hence, we can reduce $\evoII$ to
\begin{align}
	\evoII([g^t,\tilde{\tilde{\mathbf{p}}}^t])
	&=\left[{}_1e(g^t,p^t_1),{}_2e(g,p^t_1)\circ {}_2e(g,p^t_2)\circ(p^t_3,...,p_{|\mathbf{p}^{t} |}^{t})\right]\\
	& =\Bigg[\Big(\mathfrak{q}^{t} ,\underbrace{\max\left( \textbf{start},\overline{\mathfrak{q}}_{1}^{t}\right)}_{\Delta \mathfrak{q}}\Big) ,\bigg(\Bigl(\mathfrak{\underbrace{\vdash }_{\mathfrak{z}} ,}\underbrace{-1}_{\mathfrak{h}} ,\underbrace{0}_{\Delta \mathfrak{h}} ,\underbrace{-1}_{\mathfrak{o}} ,\underbrace{-1}_{\Delta \mathfrak{o}} ,\underbrace{0}_{\mathfrak{a}}\Bigr)\bigg) \notag \\
	& \ \quad \circ\bigg(\Bigl(\underbrace{x_{2}^{t}}_{\mathfrak{z}} ,\underbrace{1}_{\mathfrak{h}} ,\underbrace{0}_{\Delta \mathfrak{h}} ,\underbrace{1}_{\mathfrak{o}} ,\underbrace{-1}_{\Delta \mathfrak{o}} ,\underbrace{0}_{\mathfrak{a}}\Bigr)\bigg) \circ \left( p_{3}^{t} ,...,p_{|\mathbf{p}^{t} |}^{t}\right)\Bigg]\\
	& =\Bigg[ \big(\mathfrak{q}^{t} ,\underbrace{\overline{\mathfrak{q}}_{1}^{t}}_{\Delta \mathfrak{q}} \big),\bigg(\Bigl(\mathfrak{\underbrace{\vdash }_{\mathfrak{z}} ,}\underbrace{-1}_{\mathfrak{h}} ,\underbrace{0}_{\Delta \mathfrak{h}} ,\underbrace{-1}_{\mathfrak{o}} ,\underbrace{-1}_{\Delta \mathfrak{o}} ,\underbrace{0}_{\mathfrak{a}}\Bigr)\bigg) \notag\\
	& \ \quad \circ\bigg(\Bigl(\underbrace{x_{2}^{t}}_{\mathfrak{z}} ,\underbrace{1}_{\mathfrak{h}} ,\underbrace{0}_{\Delta \mathfrak{h}} ,\underbrace{1}_{\mathfrak{o}} ,\underbrace{-1}_{\Delta \mathfrak{o}} ,\underbrace{0}_{\mathfrak{a}}\Bigr)\bigg) \circ \left( p_{3}^{t} ,...,p_{|\mathbf{p}^{t} |}^{t}\right)\Bigg]
\end{align}
The last step for this case is the evolve function of the global variable.
\begin{equation}
	\overset{\circ }{e}\Big(\big(\mathfrak{q}^{t} ,\underbrace{\overline{\mathfrak{q}}_{1}^{t}}_{\Delta \mathfrak{q}}\big)\Big) :=\Big(\underbrace{\overline{\mathfrak{q}}_{1}^{t}}_{\mathfrak{q}} ,\underbrace{\textbf{start}}_{\Delta \mathfrak{q}}\Big) =\Big(\underbrace{q^{t+1}}_{\mathfrak{q}} ,\underbrace{\textbf{start}}_{\Delta \mathfrak{q}}\Big)
\end{equation}
Hence,
\begin{align}
	s\left(\left[ g^{t} ,\mathbf{p}^{t}\right]\right) & =\Bigg[\Big(\underbrace{q^{t+1}}_{\mathfrak{q}} ,\underbrace{\textbf{start}}_{\Delta \mathfrak{q}}\Big) ,\bigg(\Bigl(\mathfrak{\underbrace{\vdash }_{\mathfrak{z}} ,}\underbrace{-1}_{\mathfrak{h}} ,\underbrace{0}_{\Delta \mathfrak{h}} ,\underbrace{-1}_{\mathfrak{o}} ,\underbrace{-1}_{\Delta \mathfrak{o}} ,\underbrace{0}_{\mathfrak{a}}\Bigr)\bigg) \notag \\
	& \ \quad\circ \bigg(\Bigl(\underbrace{x_{2}^{t}}_{\mathfrak{z}} ,\underbrace{1}_{\mathfrak{h}} ,\underbrace{0}_{\Delta \mathfrak{h}} ,\underbrace{1}_{\mathfrak{o}} ,\underbrace{-1}_{\Delta \mathfrak{o}} ,\underbrace{0}_{\mathfrak{a}}\Bigr)\bigg) \circ \left( p_{3}^{t} ,...,p_{|\mathbf{p}^{t} |}^{t}\right)\Bigg]\\
	& =\left[ g^{t+1} ,\mathbf{p}^{t+1}\right]
\end{align}

We need to prove that the criteria (\ref{eq:TMcomplex:inductionStepConditions}) are true for $\left[ g^{t+1} ,\mathbf{p}^{t+1}\right]$.
\begin{equation}
	\begin{array}{ l }
		\psi ^{-1}\left(\left[ g^{t+1} ,\mathbf{p}^{t+1}\right]\right)\\
		=\left( q^{t+1} ,\vdash x_{2}^{t} \cdots x_{|\mathbf{p}^{t} |}^{t} \blank^{\omega } ,2\right)\\
		=\left( q^{t+1} ,\mathbf{x}^{t+1} \blank^{\omega } ,n^{t+1}\right)\\
		=\underline{\alpha ^{t+1}}
	\end{array}
\end{equation}

The rest of the criteria can be read from the calculated $\left[ g^{t+1},\mathbf{p}^{t+1}\right]$
\begin{align}
	&\mathfrak{h}_{n^{t+1}}^{t+1} =\mathfrak{h}_{2}^{t+1}=\underline{1},\\ 
	&\mathfrak{h}_{n^{t+1} -\mathfrak{o}_{n^{t+1}}^{t+1}}^{t+1} =
	\mathfrak{h}_{2 -\mathfrak{o}_{2}^{t+1}}^{t+1}=\mathfrak{h}_{2 -1}^{t+1} =\underline{  -1},\\
	&\forall j\in \left\{1,...,|\mathbf{p}^{t+1} |\right\} \backslash \Big\{ \underbrace{n^{t+1}}_{=2} ,\underbrace{n^{t+1} -\mathfrak{o}_{n^{t+1}}^{t+1}}_{=1}\Big\} :\ \mathfrak{h}_{j}^{t+1} =0,\\
	&	 \forall j\in \left\{1,...,|\mathbf{p}^{t+1} |\right\} :\Delta \mathfrak{h}_{j}^{t+1} =0\ \land \Delta \mathfrak{o}_{j}^{t+1} =-1\land \mathfrak{a}_{j}^{t+1} =0,\\
	& \Delta q^{t+1} =\textbf{start}.
\end{align}

\sloppy Second, we prove it for the case 
$\mathfrak{h}^t_{n^t}=1$, $\mathfrak{h}^t_{n^t-1}=-1$, and $\overline{\mathfrak{d}}^t_{n^t}=1$
where $n^t~\in~\{2,...,|\mathbf{p}^t|-1 \}$. Using the criteria (\ref{eq:TMcomplex:inductionStepConditions}) lead to
\begin{equation}\label{eq:TMcomplex:secondcase:PMstate}
	\left[ g^{t} ,\mathbf{p}^{t}\right] =\left[ (\underbrace{q^{t}}_{\mathfrak{q}} ,\underbrace{\textbf{start}}_{\Delta \mathfrak{q}} ),\left(\begin{array}{ c c c c c c }
		(\underbrace{x_{1}^{t}}_{\mathfrak{z}}\mathfrak{,} & \underbrace{0}_{\mathfrak{h}} , & \underbrace{0}_{\Delta \mathfrak{h}} , & \underbrace{-1}_{\mathfrak{o}} , & \underbrace{-1}_{\Delta \mathfrak{o}} , & \underbrace{0}_{\mathfrak{a}})\\
		\vdots  &  &  &  &  & \\
		( x_{n^t-2}^{t} , & 0, & 0, & -1, & -1, & 0)\\
		( x_{n^t-1}^{t} , & -1, & 0, & -1, & -1, & 0)\\
		( x_{n^t} , & 1, & 0, & 1, & -1, & 0)\\
		( x_{n^t+1}^{t} , & 0, & 0, & -1, & -1, & 0)\\
		\vdots  &  &  &  &  & \\
		( x_{|\mathbf{x}^{t} |}^{t} & 0, & 0, & -1, & -1, & 0)
	\end{array}\right)^{\mathbf{T}}\right]
\end{equation}

We use again
\begin{equation}
	\delta(\mathfrak{q}^t,\mathfrak{z}^t_{n^t})=:(\overline{\mathfrak{q}}^t,\overline{\mathfrak{z}}^t_{n^t},\overline{\mathfrak{d}}^t_{n^t}).
\end{equation}

The particle $p_{n^t}$ is the only particle with $\mathfrak{h}^t_{n^t}=1$ and $p_{n^t-1}$ with $\mathfrak{h}^t_{n^t-1}=-1$. Since $\overline{\mathfrak{d}}^t_{n^t}=1$ and $\overline{\mathfrak{o}}^t_{n^t}=1$ the interact function $i$ is the identity except for $i(g,p_{n^t},p_{n^t+1})$ and $i(g,p_{n^t+1},p_{n^t})$. Therefore,
\begin{align}
	\tilde{\mathbf{p}}^t
	&:=\intIII([g^t,\mathbf{p}^t])\\
	& =\biggl( p_{1}^{t} ,...,p_{n^t-1}^{t} , \Bigl(\underbrace{x_{n^t}^{t}}_{\mathfrak{z}}\underbrace{1}_{\mathfrak{h}} ,\underbrace{0}_{\Delta \mathfrak{h}} ,\underbrace{1}_{\mathfrak{o}} ,\underbrace{-1}_{\Delta \mathfrak{o}} ,\underbrace{1}_{\mathfrak{a}}\Bigr) ,\notag\\
	& \quad \quad \Bigl(\underbrace{x_{n^t+1}^{t}}_{\mathfrak{z}}\underbrace{0}_{\mathfrak{h}} ,\underbrace{1}_{\Delta \mathfrak{h}} ,\underbrace{-1}_{\mathfrak{o}} ,\underbrace{\max( -1,1)}_{\Delta \mathfrak{o}} ,\underbrace{0}_{\mathfrak{a}}\Bigr) ,p_{n^t+2}^{t} ,...,p_{|\mathbf{p}^{t} |}^{t}\biggr)\\
	& =\biggl( p_{1}^{t} ,...,p_{n^t-1}^{t} , \Bigl(\underbrace{x_{n^t}^{t}}_{\mathfrak{z}}\underbrace{1}_{\mathfrak{h}} ,\underbrace{0}_{\Delta \mathfrak{h}} ,\underbrace{1}_{\mathfrak{o}} ,\underbrace{-1}_{\Delta \mathfrak{o}} ,\underbrace{1}_{\mathfrak{a}}\Bigr) ,\notag\\
	& \quad \quad \Bigl(\underbrace{x_{n^t+1}^{t}}_{\mathfrak{z}}\underbrace{0}_{\mathfrak{h}} ,\underbrace{1}_{\Delta \mathfrak{h}} ,\underbrace{-1}_{\mathfrak{o}} ,\underbrace{1}_{\Delta \mathfrak{o}} ,\underbrace{0}_{\mathfrak{a}}\Bigr) ,p_{n^t+2}^{t} ,...,p_{|\mathbf{p}^{t} |}^{t}\biggr).
\end{align}
The next is the evolution step. The evolve function $e$ is the identity for all particles except if $\mathfrak{h}=1$ this is only true for $\tilde p^t_{n^t}$ or if $\Delta\mathfrak{h}=1$, only true for $\tilde p^t_{n^t+1}$, or if $\mathfrak{h}=-1$ only true for $p^t_{n^t-1}$. The evolve function $e$ is the identity for the global variable except if $\mathfrak{h}=1$.

Hence, we can reduce $\evoII$ to
\begin{align}
	\evoII([g^{t} ,\tilde{\mathbf{p}}^{t} ]) & =[_{1} e(g^{t} ,p_{1}^{t} ),(p_{1}^{t} ,...,p_{n^t-2}^{t} )\circ {}_{2} e(g, p_{n^t-1}^{t} )\notag\\
	& \quad \quad \circ {}_{2} e(g,\tilde p_{n^t}^{t} )\circ {}_{2} e(g,\tilde p_{n^t+1}^{t} )\circ (p_{n^t+2}^{t} ,...,p_{|\mathbf{p}^{t} |}^{t} )]\\
	& =\Biggl[\Bigl(\mathfrak{q}^{t} ,\underbrace{\max\left( \textbf{start},\overline{\mathfrak{q}}_{n^t}^{t}\right)}_{\Delta \mathfrak{q}}\Bigr) ,\ (p_{1}^{t} ,...,p_{n^t-2}^{t} )\notag\\
	& \quad \circ \biggl(\Bigl(\underbrace{x_{n^t-1}^{t}}_{\mathfrak{z}}\underbrace{0}_{\mathfrak{h}} ,\underbrace{0}_{\Delta \mathfrak{h}} ,\underbrace{-1}_{\mathfrak{o}} ,\underbrace{-1}_{\Delta \mathfrak{o}} ,\underbrace{0}_{\mathfrak{a}}\Bigr)\biggr)\notag\\
	& \quad \circ \left(\Bigl(\underbrace{\overline{\mathfrak{z}}_{n^t}^{t}}_{\mathfrak{z}}\underbrace{-1}_{\mathfrak{h}} ,\underbrace{0}_{\Delta \mathfrak{h}} ,\underbrace{-1}_{\mathfrak{o}} ,\underbrace{-1}_{\Delta \mathfrak{o}} ,\underbrace{0}_{\mathfrak{a}}\Bigr)\right)\notag\\
	& \quad \circ \biggl(\Bigl(\underbrace{x_{n^t+1}^{t}}_{\mathfrak{z}}\underbrace{1}_{\mathfrak{h}} ,\underbrace{0}_{\Delta \mathfrak{h}} ,\underbrace{1}_{\mathfrak{o}} ,\underbrace{-1}_{\Delta \mathfrak{o}} ,\underbrace{0}_{\mathfrak{a}}\Bigr)\biggr) \circ \left( p_{n^t+2}^{t} ,...,p_{|\mathbf{p}^{t} |}^{t}\right)\Biggr].
\end{align}

The last step for this case is the evolve function of the global variable.
\begin{equation}
	\overset{\circ }{e}\Big(\big(\mathfrak{q}^{t} ,\underbrace{\overline{\mathfrak{q}}_{n^t}^{t}}_{\Delta \mathfrak{q}}\big)\Big) :=\Big(\underbrace{\overline{\mathfrak{q}}_{n^t}^{t}}_{\mathfrak{q}} ,\underbrace{\textbf{start}}_{\Delta \mathfrak{q}}\Big) =\Big(\underbrace{q^{t+1}}_{\mathfrak{q}} ,\underbrace{\textbf{start}}_{\Delta \mathfrak{q}}\Big)
\end{equation}

Hence,
\begin{align}
	s\left(\left[ g^{t} ,\mathbf{p}^{t}\right]\right) 
	&=\left[ g^{t+1} ,\mathbf{p}^{t+1}\right]\\
	& =\Bigg[\Big(\underbrace{q^{t+1}}_{\mathfrak{q}} ,\underbrace{\textbf{start}}_{\Delta \mathfrak{q}}\Big),\ (p_{1}^{t} ,...,p_{n^t-2}^{t} )\notag\\
	& \quad \circ \biggl(\Bigl(\underbrace{x_{n^t-1}^{t}}_{\mathfrak{z}}\underbrace{0}_{\mathfrak{h}} ,\underbrace{0}_{\Delta \mathfrak{h}} ,\underbrace{-1}_{\mathfrak{o}} ,\underbrace{-1}_{\Delta \mathfrak{o}} ,\underbrace{0}_{\mathfrak{a}}\Bigr)\biggr)\notag\\
	& \quad \circ \left(\Bigl(\underbrace{\overline{\mathfrak{z}}_{n^t}^{t}}_{\mathfrak{z}}\underbrace{-1}_{\mathfrak{h}} ,\underbrace{0}_{\Delta \mathfrak{h}} ,\underbrace{-1}_{\mathfrak{o}} ,\underbrace{-1}_{\Delta \mathfrak{o}} ,\underbrace{0}_{\mathfrak{a}}\Bigr)\right)\notag\\
	& \quad \circ \biggl(\Bigl(\underbrace{x_{n^t+1}^{t}}_{\mathfrak{z}}\underbrace{1}_{\mathfrak{h}} ,\underbrace{0}_{\Delta \mathfrak{h}} ,\underbrace{1}_{\mathfrak{o}} ,\underbrace{-1}_{\Delta \mathfrak{o}} ,\underbrace{0}_{\mathfrak{a}}\Bigr)\biggr) \circ \left( p_{n^t+2}^{t} ,...,p_{|\mathbf{p}^{t} |}^{t}\right)\Biggr]
\end{align}

We need to prove again that the criteria (\ref{eq:TMcomplex:inductionStepConditions}) is true for $\left[ g^{t+1} ,\mathbf{p}^{t+1}\right]$.
\begin{equation}
	\begin{array}{ l }
		\psi ^{-1}\left(\left[ g^{t+1} ,\mathbf{p}^{t+1}\right]\right)\\
		=\left( q^{t+1} ,x_{1}^{t} \cdots x_{n^{t} -1}^{t}\overline{\mathfrak{z}}_{n^{t}}^{t} x_{n^{t} +1}^{t} \cdots x_{|\mathbf{p}^{t} |}^{t} \blank^{\omega } ,n^{t} +1\right)\\
		=\left( q^{t+1} ,\mathbf{x}^{t+1} \blank^{\omega } ,n^{t+1}\right)\\
		=\underline{\alpha ^{t+1}}
	\end{array}
\end{equation}

The rest of the criteria can be read from the calculated $\left[ g^{t+1},\mathbf{p}^{t+1}\right]$
\begin{align}
	& \mathfrak{h}_{n^{t+1}}^{t+1} =\mathfrak{h}_{n^{t} +1}^{t+1} =1 ,\\
	& \mathfrak{h}_{n^{t+1} -\mathfrak{o}_{n^{t+1}}^{t+1}}^{t+1} =\mathfrak{h}_{n^{t} +1-\mathfrak{o}_{n^{t} +1}^{t+1}}^{t+1} =\mathfrak{h}_{n^{t} +1-1}^{t+1} =-1,\\
	& \forall j\in \left\{1,...,|\mathbf{p}^{t+1} |\right\} \backslash \Bigl\{\underbrace{n^{t+1}}_{=n^{t} +1} ,\underbrace{n^{t+1} -\mathfrak{o}_{n^{t+1}}^{t+1}}_{=n^{t}}\Bigr\} :\ \mathfrak{h}_{j}^{t+1} =0,\\
	& \forall j\in \left\{1,...,|\mathbf{p}^{t+1} |\right\} :\Delta \mathfrak{h}_{j}^{t+1} =0\ \land \Delta \mathfrak{o}_{j}^{t+1} =-1\land \mathfrak{a}_{j}^{t+1} =0,\\
	& \Delta q^{t+1} =\textbf{start}.
\end{align}

Third, we prove it for the case 
$\mathfrak{h}^t_{n^t}=1$, $\mathfrak{h}^t_{n^t-1}=-1$, and $\delta(\mathfrak{q}^t,\mathfrak{z}^t_{n^t})=:(\overline{\mathfrak{q}}^t,\overline{\mathfrak{z}}^t_{n^t},\overline{\mathfrak{d}}^t_{n^t})=(q^{t+1},\overline{\mathfrak{z}}^t_{n^t},-1)$
where $n^t\in \{2,...,|\mathbf{p}^t| \}$. This results in the same state $[g^t,\mathbf{p}^t]$ as in  (\ref{eq:TMcomplex:secondcase:PMstate}).

The particle $p_{n^t}$ is the only particle with $\mathfrak{h}^t_{n^t}=1$ and $p_{n^t-1}$ with $\mathfrak{h}^t_{n^t-1}=-1$. Since $\overline{\mathfrak{d}}^t_{n^t}=-1$ and $\overline{\mathfrak{o}}^t_{n^t}=1$, the interact function is the identity except for $i(g,p_{n^t-1},p_{n^t})$ and  $i(g,p_{n^t},p_{n^t-1})$. Therefore,
\begin{align}
	\tilde{\mathbf{p}}^t
	&:=\intIII([g^t,\mathbf{p}^t])\\
	& =\biggl( p_{1}^{t} ,...,p_{n^{t} -2}^{t} ,\Bigl(\underbrace{x_{n^{t} -1}^{t}}_{\mathfrak{z}}\underbrace{-1}_{\mathfrak{h}} ,\underbrace{1}_{\Delta \mathfrak{h}} ,\underbrace{-1}_{\mathfrak{o}} ,\underbrace{\max( -1,-1)}_{\Delta \mathfrak{o}} ,\underbrace{0}_{\mathfrak{a}}\Bigr) ,\notag\\
	& \quad \quad \Bigl(\underbrace{x_{n^{t}}^{t}}_{\mathfrak{z}}\underbrace{1}_{\mathfrak{h}} ,\underbrace{0}_{\Delta \mathfrak{h}} ,\underbrace{1}_{\mathfrak{o}} ,\underbrace{-1}_{\Delta \mathfrak{o}} ,\underbrace{1}_{\mathfrak{a}}\Bigr) ,p_{n^{t} +1}^{t} ,...,p_{|\mathbf{p}^{t} |}^{t}\biggr)\\
	& =\biggl( p_{1}^{t} ,...,p_{n^{t} -2}^{t} ,\Bigl(\underbrace{x_{n^{t} -1}^{t}}_{\mathfrak{z}}\underbrace{-1}_{\mathfrak{h}} ,\underbrace{1}_{\Delta \mathfrak{h}} ,\underbrace{-1}_{\mathfrak{o}} ,\underbrace{-1}_{\Delta \mathfrak{o}} ,\underbrace{0}_{\mathfrak{a}}\Bigr) ,\notag\\
	& \quad \quad \Bigl(\underbrace{x_{n^{t}}^{t}}_{\mathfrak{z}}\underbrace{1}_{\mathfrak{h}} ,\underbrace{0}_{\Delta \mathfrak{h}} ,\underbrace{1}_{\mathfrak{o}} ,\underbrace{-1}_{\Delta \mathfrak{o}} ,\underbrace{1}_{\mathfrak{a}}\Bigr) ,p_{n^{t} +1}^{t} ,...,p_{|\mathbf{p}^{t} |}^{t}\biggr)		.
\end{align}

The next is the evolution step. The evolve function $e$ is the identity for all particles except if $\mathfrak{h}=1$ this is only true for $\tilde p^t_{n^t}$ or if $\Delta\mathfrak{h}=1$,only true for $\tilde p^t_{n^t-1}$, or if $\mathfrak{h}=-1$ only true for $\tilde p^t_{n^t-1}$, too. The evolve function $e$ is the identity for the global variable except if $\mathfrak{h}=1$. Hence, we can reduce $\evoII$ to
\begin{align}
	\evoII([g^{t} ,\tilde{\mathbf{p}}^{t} ]) & =[_{1} e(g^{t} ,p_{1}^{t} ),(p_{1}^{t} ,...,p_{n^{t} -2}^{t} )\circ {}_{2} e(g, \tilde p_{n^{t} -1}^{t} )\notag\\
	& \ \ \ \ \circ {}_{2} e(g,\tilde p_{n^{t}}^{t} )\circ (p_{n^{t} +1}^{t} ,...,p_{|\mathbf{p}^{t} |}^{t} )]\\
	& =\Biggl[\Bigl(\mathfrak{q}^{t} ,\underbrace{\max\left( \textbf{start},\overline{\mathfrak{q}}_{n^{t}}^{t}\right)}_{\Delta \mathfrak{q}}\Bigr) ,\ (p_{1}^{t} ,...,p_{n^{t} -2}^{t} )\notag\\
	& \ \ \circ \biggl(\Bigl(\underbrace{x_{n^{t} -1}^{t}}_{\mathfrak{z}}\underbrace{1}_{\mathfrak{h}} ,\underbrace{0}_{\Delta \mathfrak{h}} ,\underbrace{-1}_{\mathfrak{o}} ,\underbrace{-1}_{\Delta \mathfrak{o}} ,\underbrace{0}_{\mathfrak{a}}\Bigr)\biggr)\notag\\
	& \ \ \circ \left(\Bigl(\underbrace{\overline{\mathfrak{z}}_{n^{t}}^{t}}_{\mathfrak{z}}\underbrace{-1}_{\mathfrak{h}} ,\underbrace{0}_{\Delta \mathfrak{h}} ,\underbrace{-1}_{\mathfrak{o}} ,\underbrace{-1}_{\Delta \mathfrak{o}} ,\underbrace{0}_{\mathfrak{a}}\Bigr)\right) \circ \left( p_{n^{t} +1}^{t} ,...,p_{|\mathbf{p}^{t} |}^{t}\right)\Biggr].
\end{align}

The last step for this case is the evolve function of the global variable.
\begin{equation}
	\overset{\circ }{e}\Big(\big(\mathfrak{q}^{t} ,\underbrace{\overline{\mathfrak{q}}_{n^t}^{t}}_{\Delta \mathfrak{q}}\big)\Big) :=\Big(\underbrace{\overline{\mathfrak{q}}_{n^t}^{t}}_{\mathfrak{q}} ,\underbrace{\textbf{start}}_{\Delta \mathfrak{q}}\Big) =\Big(\underbrace{q^{t+1}}_{\mathfrak{q}} ,\underbrace{\textbf{start}}_{\Delta \mathfrak{q}}\Big)
\end{equation}

Hence,
\begin{align}
	s\left(\left[ g^{t} ,\mathbf{p}^{t}\right]\right)
	&=\left[ g^{t+1} ,\mathbf{p}^{t+1}\right]\\
	& =\Bigg[\Big(\underbrace{q^{t+1}}_{\mathfrak{q}} ,\underbrace{\textbf{start}}_{\Delta \mathfrak{q}}\Big),\ (p_{1}^{t} ,...,p_{n^t-2}^{t} )\notag\\
	& \ \ \circ \biggl(\Bigl(\underbrace{x_{n^{t} -1}^{t}}_{\mathfrak{z}}\underbrace{1}_{\mathfrak{h}} ,\underbrace{0}_{\Delta \mathfrak{h}} ,\underbrace{-1}_{\mathfrak{o}} ,\underbrace{-1}_{\Delta \mathfrak{o}} ,\underbrace{0}_{\mathfrak{a}}\Bigr)\biggr)\notag\\
	& \ \ \circ \left(\Bigl(\underbrace{\overline{\mathfrak{z}}_{n^{t}}^{t}}_{\mathfrak{z}}\underbrace{-1}_{\mathfrak{h}} ,\underbrace{0}_{\Delta \mathfrak{h}} ,\underbrace{-1}_{\mathfrak{o}} ,\underbrace{-1}_{\Delta \mathfrak{o}} ,\underbrace{0}_{\mathfrak{a}}\Bigr)\right) \circ \left( p_{n^{t} +1}^{t} ,...,p_{|\mathbf{p}^{t} |}^{t}\right)\Biggr].
\end{align}

We need to prove again that the criteria (\ref{eq:TMcomplex:inductionStepConditions}) are true for $\left[ g^{t+1} ,\mathbf{p}^{t+1}\right]$.
\begin{equation}
	\begin{array}{ l }
		\psi ^{-1}\left(\left[ g^{t+1} ,\mathbf{p}^{t+1}\right]\right)\\
		=\left( q^{t+1} ,x_{1}^{t} \cdots x_{n^{t} -1}^{t}\overline{\mathfrak{z}}_{n^{t}}^{t} x_{n^{t} +1}^{t} \cdots x_{|\mathbf{p}^{t} |}^{t} \blank^{\omega } ,n^{t} -1\right)\\
		=\left( q^{t+1} ,\mathbf{x}^{t+1} \blank^{\omega } ,n^{t+1}\right)\\
		=\underline{\alpha ^{t+1}}
	\end{array}
\end{equation}

The rest of the criteria can be read from the calculated $\left[ g^{t+1},\mathbf{p}^{t+1}\right]$

\begin{align}
	& \mathfrak{h}_{n^{t+1}}^{t+1} =\mathfrak{h}_{n^{t} -1}^{t+1} =1,\\
	& \mathfrak{h}_{n^{t+1} -\mathfrak{o}_{n^{t+1}}^{t+1}}^{t+1} =\mathfrak{h}_{n^{t} -1-\mathfrak{o}_{n^{t} -1}^{t+1}}^{t+1} =\mathfrak{h}_{n^{t} -1-( -1)}^{t+1} =-1,\\
	& \forall j\in \left\{1,...,|\mathbf{p}^{t+1} |\right\} \backslash \Bigl\{\underbrace{n^{t+1}}_{=n^{t} -1} ,\underbrace{n^{t+1} -\mathfrak{o}_{n^{t+1}}^{t+1}}_{=n^{t}}\Bigr\} :\ \mathfrak{h}_{j}^{t+1} =0,\\
	& \forall j\in \left\{1,...,|\mathbf{p}^{t+1} |\right\} :\Delta \mathfrak{h}_{j}^{t+1} =0\ \land \Delta \mathfrak{o}_{j}^{t+1} =-1\land \mathfrak{a}_{j}^{t+1} =0,\\
	& \Delta q^{t+1} =\textbf{start}.
\end{align}

Fourth, we prove it for the case 
$\mathfrak{h}^t_{n^t}=1$, $\mathfrak{h}^t_{n^t+1}=-1$, and $\delta(\mathfrak{q}^t,\mathfrak{z}^t_{n^t})=:(\overline{\mathfrak{q}}^t,\overline{\mathfrak{z}}^t_{n^t},\overline{\mathfrak{d}}^t_{n^t})=(q^{t+1},\overline{\mathfrak{z}}^t_{n^t},1)$
where $n^t\in \{2,...,|\mathbf{p}^t|-1 \}$. Using the criteria (\ref{eq:TMcomplex:inductionStepConditions}) lead to
\begin{equation}\label{eq:TMcomplex:fourthcase:PMstate}
	\left[ g^{t} ,\mathbf{p}^{t}\right] =\left[ (\underbrace{q^{t}}_{\mathfrak{q}} ,\underbrace{\textbf{start}}_{\Delta \mathfrak{q}} ),\left(\begin{array}{ c c c c c c }
		(\underbrace{x_{1}^{t}}_{\mathfrak{z}}\mathfrak{,} & \underbrace{0}_{\mathfrak{h}} , & \underbrace{0}_{\Delta \mathfrak{h}} , & \underbrace{-1}_{\mathfrak{o}} , & \underbrace{-1}_{\Delta \mathfrak{o}} , & \underbrace{0}_{\mathfrak{a}} )\\
		\vdots  &  &  &  &  & \\
		(x_{n^{t} -1}^{t} , & 0, & 0, & -1, & -1, & 0)\\
		(x_{n^{t}} , & 1, & 0, & -1, & -1, & 0)\\
		(x_{n^{t} +1}^{t} , & -1, & 0, & -1, & -1, & 0)\\
		(x_{n^{t} +2}^{t} , & 0, & 0, & -1, & -1, & 0)\\
		\vdots  &  &  &  &  & \\
		(x_{|\mathbf{x}^{t} |}^{t} & 0, & 0, & -1, & -1, & 0)
	\end{array}\right)^{\mathbf{T}}\right]
\end{equation}

The particle $p_{n^t}$ is the only particle with $\mathfrak{h}^t_{n^t}=1$ and $p_{n^t+1}$ with $\mathfrak{h}^t_{n^t+1}=-1$. Since $\overline{\mathfrak{d}}^t_{n^t}=1$ and $\overline{\mathfrak{o}}^t_{n^t}=-1$ the interact function is the identity except for  $i(g,p_{n^t},p_{n^t+1})$ and $i(g,p_{n^t+1},p_{n^t})$. Therefore,
\begin{align}
	\tilde{\mathbf{p}}^t
	&:=\intIII([g^t,\mathbf{p}^t])\\
	& =\biggl( p_{1}^{t} ,...,p_{n^{t} -1}^{t} ,\Bigl(\underbrace{x_{n^{t}}^{t}}_{\mathfrak{z}}\underbrace{1}_{\mathfrak{h}} ,\underbrace{0}_{\Delta \mathfrak{h}} ,\underbrace{-1}_{\mathfrak{o}} ,\underbrace{-1}_{\Delta \mathfrak{o}} ,\underbrace{1}_{\mathfrak{a}}\Bigr) ,\notag\\
	& \quad \quad \Bigl(\underbrace{x_{n^{t} +1}^{t}}_{\mathfrak{z}}\underbrace{-1}_{\mathfrak{h}} ,\underbrace{1}_{\Delta \mathfrak{h}} ,\underbrace{-1}_{\mathfrak{o}} ,\underbrace{\max( -1,1)}_{\Delta \mathfrak{o}} ,\underbrace{0}_{\mathfrak{a}}\Bigr) ,p_{n^{t} +2}^{t} ,...,p_{|\mathbf{p}^{t} |}^{t}\biggr).
\end{align}

The next is the evolution step. The evolve function $e$ is the identity for all particles except if $\mathfrak{h}=1$ this is only true for $\tilde p^t_{n^t}$ or if $\Delta\mathfrak{h}=1$, only true for $\tilde p^t_{n^t+1}$, or if $\mathfrak{h}=-1$ only true for $\tilde p^t_{n^t+1}$, too. $e$ is the identity for the global variable except if $\mathfrak{h}=1$. Hence, we can reduce $\evoII$ to
\begin{align}
	\evoII([g^{t} ,\tilde{\mathbf{p}}^{t} ]) & =[_{1} e(g^{t} ,p_{1}^{t} ),(p_{1}^{t} ,...,p_{n^{t} -1}^{t} )\circ _{2} e(g,\tilde{p}_{n^{t}}^{t} )\notag\\
	& \ \ \ \ \circ _{2} e(g,\tilde{p}_{n^{t} +1}^{t} )\circ (p_{n^{t} +2}^{t} ,...,p_{|\mathbf{p}^{t} |}^{t} )]\\
	& =\Biggl[\Bigl(\mathfrak{q}^{t} ,\underbrace{\max\left( \textbf{start},\overline{\mathfrak{q}}_{n^{t}}^{t}\right)}_{\Delta \mathfrak{q}}\Bigr) ,\ (p_{1}^{t} ,...,p_{n^{t} -1}^{t} )\notag\\
	& \ \ \circ \left(\Bigl(\underbrace{\overline{\mathfrak{z}}_{n^{t}}^{t}}_{\mathfrak{z}}\underbrace{-1}_{\mathfrak{h}} ,\underbrace{0}_{\Delta \mathfrak{h}} ,\underbrace{-1}_{\mathfrak{o}} ,\underbrace{-1}_{\Delta \mathfrak{o}} ,\underbrace{0}_{\mathfrak{a}}\Bigr)\right)\notag\\
	& \ \ \circ \biggl(\Bigl(\underbrace{x_{n^{t} +1}^{t}}_{\mathfrak{z}}\underbrace{1}_{\mathfrak{h}} ,\underbrace{0}_{\Delta \mathfrak{h}} ,\underbrace{1}_{\mathfrak{o}} ,\underbrace{-1}_{\Delta \mathfrak{o}} ,\underbrace{0}_{\mathfrak{a}}\Bigr)\biggr) \circ \left( p_{n^{t} +2}^{t} ,...,p_{|\mathbf{p}^{t} |}^{t}\right)\Biggr] .
\end{align}

The last step for this case is the evolve function of the global variable.
\begin{equation}
	\overset{\circ }{e}\Big(\big(\mathfrak{q}^{t} ,\underbrace{\overline{\mathfrak{q}}_{n^t}^{t}}_{\Delta \mathfrak{q}}\big)\Big) :=\Big(\underbrace{\overline{\mathfrak{q}}_{n^t}^{t}}_{\mathfrak{q}} ,\underbrace{\textbf{start}}_{\Delta \mathfrak{q}}\Big) =\Big(\underbrace{q^{t+1}}_{\mathfrak{q}} ,\underbrace{\textbf{start}}_{\Delta \mathfrak{q}}\Big)
\end{equation}

Hence,
\begin{align}
	s\left(\left[ g^{t} ,\mathbf{p}^{t}\right]\right) 
	&=\left[ g^{t+1} ,\mathbf{p}^{t+1}\right]\\
	& =\Bigg[\Big(\underbrace{q^{t+1}}_{\mathfrak{q}} ,\underbrace{\textbf{start}}_{\Delta \mathfrak{q}}\Big),\ (p_{1}^{t} ,...,p_{n^{t} -1}^{t} )\notag\\
	& \ \ \circ \left(\Bigl(\underbrace{\overline{\mathfrak{z}}_{n^{t}}^{t}}_{\mathfrak{z}}\underbrace{-1}_{\mathfrak{h}} ,\underbrace{0}_{\Delta \mathfrak{h}} ,\underbrace{-1}_{\mathfrak{o}} ,\underbrace{-1}_{\Delta \mathfrak{o}} ,\underbrace{0}_{\mathfrak{a}}\Bigr)\right)\notag\\
	& \ \ \circ \biggl(\Bigl(\underbrace{x_{n^{t} +1}^{t}}_{\mathfrak{z}}\underbrace{1}_{\mathfrak{h}} ,\underbrace{0}_{\Delta \mathfrak{h}} ,\underbrace{1}_{\mathfrak{o}} ,\underbrace{-1}_{\Delta \mathfrak{o}} ,\underbrace{0}_{\mathfrak{a}}\Bigr)\biggr) \circ \left( p_{n^{t} +2}^{t} ,...,p_{|\mathbf{p}^{t} |}^{t}\right)\Biggr] .
\end{align}

We need to prove again that the criteria (\ref{eq:TMcomplex:inductionStepConditions}) are true for $\left[ g^{t+1} ,\mathbf{p}^{t+1}\right]$.
\begin{equation}
	\begin{array}{ l }
		\psi ^{-1}\left(\left[ g^{t+1} ,\mathbf{p}^{t+1}\right]\right)\\
		=\left( q^{t+1} ,x_{1}^{t} \cdots x_{n^{t} -1}^{t}\overline{\mathfrak{z}}_{n^{t}}^{t} x_{n^{t} +1}^{t} \cdots x_{|\mathbf{p}^{t} |}^{t} \blank^{\omega } ,n^{t} +1\right)\\
		=\left( q^{t+1} ,\mathbf{x}^{t+1} \blank^{\omega } ,n^{t+1}\right)\\
		=\underline{\alpha ^{t+1}}
	\end{array}
\end{equation}

The rest of the criteria can be read from the calculated $\left[ g^{t+1},\mathbf{p}^{t+1}\right]$

\begin{align}
	& \mathfrak{h}_{n^{t+1}}^{t+1} =\mathfrak{h}_{n^{t} +1}^{t+1} =1 ,\\
	& \mathfrak{h}_{n^{t+1} -\mathfrak{o}_{n^{t+1}}^{t+1}}^{t+1} =\mathfrak{h}_{n^{t} +1-\mathfrak{o}_{n^{t} +1}^{t+1}}^{t+1} =\mathfrak{h}_{n^{t} +1-1}^{t+1} =-1,\\
	& \forall j\in \left\{1,...,|\mathbf{p}^{t+1} |\right\} \backslash \Bigl\{\underbrace{n^{t+1}}_{=n^{t} +1} ,\underbrace{n^{t+1} -\mathfrak{o}_{n^{t+1}}^{t+1}}_{=n^{t}}\Bigr\} :\ \mathfrak{h}_{j}^{t+1} =0,\\
	& \forall j\in \left\{1,...,|\mathbf{p}^{t+1} |\right\} :\Delta \mathfrak{h}_{j}^{t+1} =0\ \land \Delta \mathfrak{o}_{j}^{t+1} =-1\land \mathfrak{a}_{j}^{t+1} =0,\\
	& \Delta q^{t+1} =\textbf{start}.
\end{align}

Fifth, we prove it for the case 
$\mathfrak{h}^t_{n^t}=1$, $\mathfrak{h}^t_{n^t+1}=-1$, and $\delta(\mathfrak{q}^t,\mathfrak{z}^t_{n^t})=:(\overline{\mathfrak{q}}^t,\overline{\mathfrak{z}}^t_{n^t},\overline{\mathfrak{d}}^t_{n^t})=(q^{t+1},\overline{\mathfrak{z}}^t_{n^t},-1)$
where $n^t\in \{2,...,|\mathbf{p}^t|-1 \}$. This results in the same state $[g^t,\mathbf{p}^t]$ as in  (\ref{eq:TMcomplex:fourthcase:PMstate}).

The particle $p_{n^t}$ is the only particle with $\mathfrak{h}^t_{n^t}=1$ and $p_{n^t+1}$ with $\mathfrak{h}^t_{n^t+1}=-1$. Since $\overline{\mathfrak{d}}^t_{n^t}=-1$ and $\overline{\mathfrak{o}}^t_{n^t}=-1$ the interact function is the identity except for   $i(g,p_{n^t-1},p_{n^t})$ and $i(g,p_{n^t},p_{n^t-1})$. Therefore,
\begin{align}
	\tilde{\mathbf{p}}^t
	&:=\intIII([g^t,\mathbf{p}^t])\\
	&=\biggl( p_{1}^{t} ,...,p_{n^{t} -2}^{t} ,\Bigl(\underbrace{x_{n^{t} -1}^{t}}_{\mathfrak{z}}\underbrace{0}_{\mathfrak{h}} ,\underbrace{1}_{\Delta \mathfrak{h}} ,\underbrace{-1}_{\mathfrak{o}} ,\underbrace{\max( -1,-1)}_{\Delta \mathfrak{o}} ,\underbrace{0}_{\mathfrak{a}}\Bigr) ,\notag\\
	&\quad \quad \Bigl(\underbrace{x_{n^{t}}^{t}}_{\mathfrak{z}}\underbrace{1}_{\mathfrak{h}} ,\underbrace{0}_{\Delta \mathfrak{h}} ,\underbrace{-1}_{\mathfrak{o}} ,\underbrace{-1}_{\Delta \mathfrak{o}} ,\underbrace{1}_{\mathfrak{a}}\Bigr) ,p_{n^{t} +1}^{t} ,...,p_{|\mathbf{p}^{t} |}^{t}\biggr)
\end{align}

The next is the evolution step. The evolve function $e$ is the identity for all particles except if $\mathfrak{h}=1$ this is only true for $\tilde p^t_{n^t}$ or if $\Delta\mathfrak{h}=1$, only true for $\tilde p^t_{n^t-1}$, or if $\mathfrak{h}=-1$ only true for $\tilde p^t_{n^t+1}$. $e$ is the identity for the global variable except if $\mathfrak{h}=1$. Hence, we can reduce $\evoII$ to
\begin{align}
	\evoII([g^{t} ,\tilde{\mathbf{p}}^{t} ]) & =[_{1} e(g^{t} ,p_{1}^{t} ),(p_{1}^{t} ,...,p_{n^t-2}^{t} )\circ _{2} e(g,p_{n^t-1}^{t} )\notag\\
	& \quad \quad \circ _{2} e(g,p_{n^t}^{t} )\circ _{2} e(g,p_{n^t+1}^{t} )\circ (p_{n^t+2}^{t} ,...,p_{|\mathbf{p}^{t} |}^{t} )]\\
	& =\Biggl[\Bigl(\mathfrak{q}^{t} ,\underbrace{\max\left( \textbf{start},\overline{\mathfrak{q}}_{n^t}^{t}\right)}_{\Delta \mathfrak{q}}\Bigr) ,\ (p_{1}^{t} ,...,p_{n^t-2}^{t} )\notag\\
	& \quad \circ \biggl(\Bigl(\underbrace{x_{n^t-1}^{t}}_{\mathfrak{z}}\underbrace{1}_{\mathfrak{h}} ,\underbrace{0}_{\Delta \mathfrak{h}} ,\underbrace{-1}_{\mathfrak{o}} ,\underbrace{-1}_{\Delta \mathfrak{o}} ,\underbrace{0}_{\mathfrak{a}}\Bigr)\biggr)\notag\\
	& \quad \circ \left(\Bigl(\underbrace{\overline{\mathfrak{z}}_{n^t}^{t}}_{\mathfrak{z}}\underbrace{-1}_{\mathfrak{h}} ,\underbrace{0}_{\Delta \mathfrak{h}} ,\underbrace{-1}_{\mathfrak{o}} ,\underbrace{-1}_{\Delta \mathfrak{o}} ,\underbrace{0}_{\mathfrak{a}}\Bigr)\right)\notag\\
	& \quad \circ \biggl(\Bigl(\underbrace{x_{n^t+1}^{t}}_{\mathfrak{z}}\underbrace{0}_{\mathfrak{h}} ,\underbrace{0}_{\Delta \mathfrak{h}} ,\underbrace{-1}_{\mathfrak{o}} ,\underbrace{-1}_{\Delta \mathfrak{o}} ,\underbrace{0}_{\mathfrak{a}}\Bigr)\biggr) \circ \left( p_{n^t+2}^{t} ,...,p_{|\mathbf{p}^{t} |}^{t}\right)\Biggr].
\end{align}

The last step for this case is the evolve function of the global variable.
\begin{equation}
	\overset{\circ }{e}\Big(\big(\mathfrak{q}^{t} ,\underbrace{\overline{\mathfrak{q}}_{n^t}^{t}}_{\Delta \mathfrak{q}}\big)\Big) :=\Big(\underbrace{\overline{\mathfrak{q}}_{n^t}^{t}}_{\mathfrak{q}} ,\underbrace{\textbf{start}}_{\Delta \mathfrak{q}}\Big) =\Big(\underbrace{q^{t+1}}_{\mathfrak{q}} ,\underbrace{\textbf{start}}_{\Delta \mathfrak{q}}\Big)
\end{equation}
Hence,
\begin{align}
	s\left(\left[ g^{t} ,\mathbf{p}^{t}\right]\right) 
	&=\left[ g^{t+1} ,\mathbf{p}^{t+1}\right]\\
	& =\Bigg[\Big(\underbrace{q^{t+1}}_{\mathfrak{q}} ,\underbrace{\textbf{start}}_{\Delta \mathfrak{q}}\Big),\ (p_{1}^{t} ,...,p_{n^t-2}^{t} )\notag\\
	& \quad \circ \biggl(\Bigl(\underbrace{x_{n^t-1}^{t}}_{\mathfrak{z}}\underbrace{1}_{\mathfrak{h}} ,\underbrace{0}_{\Delta \mathfrak{h}} ,\underbrace{-1}_{\mathfrak{o}} ,\underbrace{-1}_{\Delta \mathfrak{o}} ,\underbrace{0}_{\mathfrak{a}}\Bigr)\biggr)\notag\\
	& \quad \circ \left(\Bigl(\underbrace{\overline{\mathfrak{z}}_{n^t}^{t}}_{\mathfrak{z}}\underbrace{-1}_{\mathfrak{h}} ,\underbrace{0}_{\Delta \mathfrak{h}} ,\underbrace{-1}_{\mathfrak{o}} ,\underbrace{-1}_{\Delta \mathfrak{o}} ,\underbrace{0}_{\mathfrak{a}}\Bigr)\right)\notag\\
	& \quad \circ \biggl(\Bigl(\underbrace{x_{n^t+1}^{t}}_{\mathfrak{z}}\underbrace{0}_{\mathfrak{h}} ,\underbrace{0}_{\Delta \mathfrak{h}} ,\underbrace{-1}_{\mathfrak{o}} ,\underbrace{-1}_{\Delta \mathfrak{o}} ,\underbrace{0}_{\mathfrak{a}}\Bigr)\biggr) \circ \left( p_{n^t+2}^{t} ,...,p_{|\mathbf{p}^{t} |}^{t}\right)\Biggr]
\end{align}

We need to prove again that the criteria (\ref{eq:TMcomplex:inductionStepConditions}) are true for $\left[ g^{t+1} ,\mathbf{p}^{t+1}\right]$.
\begin{equation}
	\begin{array}{ l }
		\psi ^{-1}\left(\left[ g^{t+1} ,\mathbf{p}^{t+1}\right]\right)\\
		=\left( q^{t+1} ,x_{1}^{t} \cdots x_{n^{t} -1}^{t}\overline{\mathfrak{z}}_{n^{t}}^{t} x_{n^{t} +1}^{t} \cdots x_{|\mathbf{p}^{t} |}^{t} \blank^{\omega } ,n^{t} -1\right)\\
		=\left( q^{t+1} ,\mathbf{x}^{t+1} \blank^{\omega } ,n^{t+1}\right)\\
		=\underline{\alpha ^{t+1}}
	\end{array}
\end{equation}

The rest of the criteria can be read from the calculated $\left[ g^{t+1},\mathbf{p}^{t+1}\right]$
\begin{align}
	& \mathfrak{h}_{n^{t+1}}^{t+1} =\mathfrak{h}_{n^{t} -1}^{t+1} =1,\\
	& \mathfrak{h}_{n^{t+1} -\mathfrak{o}_{n^{t+1}}^{t+1}}^{t+1} =\mathfrak{h}_{n^{t} -1-\mathfrak{o}_{n^{t} -1}^{t+1}}^{t+1} =\mathfrak{h}_{n^{t} -1-( -1)}^{t+1} =-1,\\
	& \forall j\in \left\{1,...,|\mathbf{p}^{t+1} |\right\} \backslash \Bigl\{\underbrace{n^{t+1}}_{=n^{t} -1} ,\underbrace{n^{t+1} -\mathfrak{o}_{n^{t+1}}^{t+1}}_{=n^{t}}\Bigr\} :\ \mathfrak{h}_{j}^{t+1} =0,\\
	& \forall j\in \left\{1,...,|\mathbf{p}^{t+1} |\right\} :\Delta \mathfrak{h}_{j}^{t+1} =0\ \land \Delta \mathfrak{o}_{j}^{t+1} =-1\land \mathfrak{a}_{j}^{t+1} =0,\\
	& \Delta q^{t+1} =\textbf{start}.
\end{align}

Sixth and last, we prove it for the case $n^t=|\mathbf{p}^t|$ with  
$\mathfrak{h}^t_{|\mathbf{p}^t|}=1$ and
$\delta(\mathfrak{q}^t,\mathfrak{z}^t_{|\mathbf{p}^t|})=:(\overline{\mathfrak{q}}^t,\overline{\mathfrak{z}}^t_{|\mathbf{p}^t|},\overline{\mathfrak{d}}^t_{|\mathbf{p}^t|})=(q^{t+1},\overline{\mathfrak{z}}^t_{|\mathbf{p}^t|},1)$. From this follows that $\mathfrak{h}^t_{|\mathbf{p}^t|-1}=-1$ and $\mathfrak{o}^t_{|\mathbf{p}^t|}=1$.

Using the criteria (\ref{eq:TMcomplex:inductionStepConditions}) lead to
\begin{equation}\label{eq:TMcomplex:sixthcase:PMstate}
	\left[ g^{t} ,\mathbf{p}^{t}\right] =\left[ (\underbrace{q^{t}}_{\mathfrak{q}} ,\underbrace{\textbf{start}}_{\Delta \mathfrak{q}} ),\left(\begin{array}{ c c c c c c }
		(\underbrace{x_{1}^{t}}_{\mathfrak{z}}\mathfrak{,} & \underbrace{0}_{\mathfrak{h}} , & \underbrace{0}_{\Delta \mathfrak{h}} , & \underbrace{-1}_{\mathfrak{o}} , & \underbrace{-1}_{\Delta \mathfrak{o}} , & \underbrace{0}_{\mathfrak{a}} )\\
		\vdots  &  &  &  &  & \\
		(x_{|\mathbf{p}^{t} |-2}^{t} , & 0, & 0, & -1, & -1, & 0)\\
		(x_{|\mathbf{p}^{t} |-1}^{t} , & -1, & 0, & -1, & -1, & 0)\\
		(x_{|\mathbf{p}^{t} |}^{t} & 1, & 0, & 1, & -1, & 0)
	\end{array}\right)^{\mathbf{T}}\right]
\end{equation}

The particle $p_{|\mathbf{p}^t|}$ is the only particle with $\mathfrak{h}^t_{|\mathbf{p}^t|}=1$ and $p_{|\mathbf{p}^t|-1}$ with $\mathfrak{h}^t_{|\mathbf{p}^t|-1}=-1$. Since $\overline{\mathfrak{d}}^t_{n^t}=1$ and $\overline{\mathfrak{o}}^t_{n^t}=1$ the interact function is the identity for all particle. Therefore,
\begin{align}
	\tilde{\mathbf{p}}^t
	&:=\intIII([g^t,\mathbf{p}^t])\\
	&=\mathbf{p}^{t}.
\end{align}

The next is the evolution step. The evolve function $e$ is the identity for all particles except if $\mathfrak{h}=1$ this is only true for $\tilde p^t_{|\mathbf{p}^t|}$ or if $\Delta\mathfrak{h}=1$, true for no particle, or if $\mathfrak{h}=-1$ only true for $\tilde p^t_{|\mathbf{p}^t|-1}$. $e$ is the identity for the global variable except if $\mathfrak{h}=1$. The particle $p_{|\mathbf{p}^t|}$ did not interact, so $\mathfrak{a}_{|\mathbf{p}^t|}=0$. The evolve function has a special case for this, where a new particle is generated. We can write $\evoII$ as
\begin{align}
	\evoII	([g^{t} ,\tilde{\mathbf{p}}^{t} ]) & =\left[_{1} e(g^{t} ,p_{1}^{t} ),(p_{1}^{t} ,...,p_{|\mathbf{p}^{t} |-2}^{t} )\circ _{2} e(g,p_{|\mathbf{p}^{t} |-1}^{t} )\circ _{2} e(g,p_{|\mathbf{p}^{t} |}^{t} )\right]\\
	& =\Biggl[\Bigl(\mathfrak{q}^{t} ,\underbrace{\max\left( \textbf{start},\overline{\mathfrak{q}}_{|\mathbf{p}^{t} |}^{t}\right)}_{\Delta \mathfrak{q}}\Bigr) ,\ (p_{1}^{t} ,...,p_{|\mathbf{p}^{t} |-2}^{t} )\notag\\
	& \quad \circ \biggl(\Bigl(\underbrace{x_{|\mathbf{p}^{t} |-1}^{t}}_{\mathfrak{z}}\underbrace{0}_{\mathfrak{h}} ,\underbrace{0}_{\Delta \mathfrak{h}} ,\underbrace{-1}_{\mathfrak{o}} ,\underbrace{-1}_{\Delta \mathfrak{o}} ,\underbrace{0}_{\mathfrak{a}}\Bigr)\biggr)\notag\\
	& \quad \circ \biggl(\Bigl(\underbrace{\overline{\mathfrak{z}}_{|\mathbf{p}^{t} |}^{t}}_{\mathfrak{z}}\underbrace{-1}_{\mathfrak{h}} ,\underbrace{0}_{\Delta \mathfrak{h}} ,\underbrace{-1}_{\mathfrak{o}} ,\underbrace{-1}_{\Delta \mathfrak{o}} ,\underbrace{0}_{\mathfrak{a}}\Bigr) ,\Bigl(\underbrace{\_}_{\mathfrak{z}}\underbrace{1}_{\mathfrak{h}} ,\underbrace{0}_{\Delta \mathfrak{h}} ,\underbrace{1}_{\mathfrak{o}} ,\underbrace{-1}_{\Delta \mathfrak{o}} ,\underbrace{0}_{\mathfrak{a}}\Bigr)\biggr)\Biggr].
\end{align}

The last step for this case is the evolve function of the global variable.
\begin{equation}
	\overset{\circ }{e}\Big(\big(\mathfrak{q}^{t} ,\underbrace{\overline{\mathfrak{q}}_{n^t}^{t}}_{\Delta \mathfrak{q}}\big)\Big) :=\Big(\underbrace{\overline{\mathfrak{q}}_{n^t}^{t}}_{\mathfrak{q}} ,\underbrace{\textbf{start}}_{\Delta \mathfrak{q}}\Big) =\Big(\underbrace{q^{t+1}}_{\mathfrak{q}} ,\underbrace{\textbf{start}}_{\Delta \mathfrak{q}}\Big)
\end{equation}
Hence,
\begin{align}
	s\left(\left[ g^{t} ,\mathbf{p}^{t}\right]\right) 
	&=\left[ g^{t+1} ,\mathbf{p}^{t+1}\right]\\
	& =\Bigg[\Big(\underbrace{q^{t+1}}_{\mathfrak{q}} ,\underbrace{\textbf{start}}_{\Delta \mathfrak{q}}\Big),\ (p_{1}^{t} ,...,p_{|\mathbf{p}^{t} |-2}^{t} )\notag\\
	& \quad \circ \biggl(\Bigl(\underbrace{x_{|\mathbf{p}^{t} |-1}^{t}}_{\mathfrak{z}}\underbrace{0}_{\mathfrak{h}} ,\underbrace{0}_{\Delta \mathfrak{h}} ,\underbrace{-1}_{\mathfrak{o}} ,\underbrace{-1}_{\Delta \mathfrak{o}} ,\underbrace{0}_{\mathfrak{a}}\Bigr)\biggr)\notag\\
	& \quad \circ \biggl(\Bigl(\underbrace{\overline{\mathfrak{z}}_{|\mathbf{p}^{t} |}^{t}}_{\mathfrak{z}}\underbrace{-1}_{\mathfrak{h}} ,\underbrace{0}_{\Delta \mathfrak{h}} ,\underbrace{-1}_{\mathfrak{o}} ,\underbrace{-1}_{\Delta \mathfrak{o}} ,\underbrace{0}_{\mathfrak{a}}\Bigr) ,\Bigl(\underbrace{\_}_{\mathfrak{z}}\underbrace{1}_{\mathfrak{h}} ,\underbrace{0}_{\Delta \mathfrak{h}} ,\underbrace{1}_{\mathfrak{o}} ,\underbrace{-1}_{\Delta \mathfrak{o}} ,\underbrace{0}_{\mathfrak{a}}\Bigr)\biggr)\Biggr].
\end{align}

We need to prove again that the criteria (\ref{eq:TMcomplex:inductionStepConditions}) are true for $\left[ g^{t+1} ,\mathbf{p}^{t+1}\right]$.
\begin{equation}
	\begin{array}{ l }
		\psi ^{-1}\left(\left[ g^{t+1} ,\mathbf{p}^{t+1}\right]\right)\\
		=\left( q^{t+1} ,x_{1}^{t} \cdots x_{|\mathbf{p}^{t} |-1}^{t}\overline{\mathfrak{z}}_{|\mathbf{p}^{t} |}^{t}\blank \blank^{\omega } ,|\mathbf{p}^{t} |+1\right)\\
		=\left( q^{t+1} ,\mathbf{x}^{t+1} \blank^{\omega } ,n^{t+1}\right)\\
		=\underline{\alpha ^{t+1}}
	\end{array}
\end{equation}
Note that $\blank \blank^{\omega}= \blank^{\omega}$, because $\omega$ is the smallest infinite ordinal number.

The rest of the criteria can be read from the calculated $\left[ g^{t+1},\mathbf{p}^{t+1}\right]$
\begin{align}
	& \mathfrak{h}_{n^{t+1}}^{t+1} =\mathfrak{h}_{|\mathbf{p}^{t} |+1}^{t+1} =1,\\
	& \mathfrak{h}_{n^{t+1} -\mathfrak{o}_{n^{t+1}}^{t+1}}^{t+1} =\mathfrak{h}_{|\mathbf{p}^{t} |+1-\mathfrak{o}_{|\mathbf{p}^{t} |+1}^{t+1}}^{t+1} =\mathfrak{h}_{|\mathbf{p}^{t} |+1-1}^{t+1} =-1,\\
	& \forall j\in \left\{1,...,|\mathbf{p}^{t+1} |\right\} \backslash \Bigl\{\underbrace{n^{t+1}}_{=|\mathbf{p}^{t} |+1} ,\underbrace{n^{t+1} -\mathfrak{o}_{n^{t+1}}^{t+1}}_{=|\mathbf{p}^{t} |}\Bigr\} :\ \mathfrak{h}_{j}^{t+1} =0,\\
	& \forall j\in \left\{1,...,|\mathbf{p}^{t+1} |\right\} :\Delta \mathfrak{h}_{j}^{t+1} =0\ \land \Delta \mathfrak{o}_{j}^{t+1} =-1\land \mathfrak{a}_{j}^{t+1} =0,\\
	& \Delta q^{t+1} =\textbf{start}.
\end{align}
With this, we proved for all cases that the particle method can simulate each configuration transition of the Turing machine. The final part is to prove that the particle method stops if and only if the Turing machine halts. 
\begin{equation}
	f(g^t)=\top \ \leftrightarrow \ q^t\in \{accept, reject\}
\end{equation} 
This is proven by
\begin{align}
	& f\left( g^{t}\right) =\top \\
	\leftrightarrow  &\ \mathfrak{q}^{t} \in \{accept,\ reject\}\\
	\xleftrightarrow{\psi ^{-1}\left(\left[ g^{t} ,\mathbf{p}^{t}\right]\right) =\alpha ^{t}} &\ q^{t} \in \{accept,\ reject\}.
\end{align}
\end{proof}

\end{document}